%% file: JeffreysFisherRaoGaussBregmanCenters-arxivV1.tex
\documentclass[11pt]{article}

\usepackage{fullpage,graphicx,amssymb,url}
\usepackage{orcidlink}
\usepackage[boxed]{algorithm2e}
\usepackage{comment}
\usepackage{rotating}
\usepackage{listings}
\graphicspath{{./}{./Fig/}}

\newtheorem{Definition}{Definition}
\newtheorem{Remark}{Remark}
\newtheorem{Theorem}{Theorem}

\newtheorem{Property}{Property}
\newtheorem{Proposition}{Proposition}
\newenvironment{Proof}{\paragraph{Proof:}}{\hfill$\square$}

\def\myinclude#1{{\typeout{Frank: Including file '#1'} \let\clearpage\relax\include{#1}}}

\input{macros-SBDCentroids.tex}

\begin{document}

\title{Fast proxy centers for Jeffreys centroids:  The Jeffreys-Fisher-Rao  and the inductive Gauss-Bregman  centers}

\author{Frank Nielsen\\ \ \\ Sony Computer Science Laboratories Inc.\\ Tokyo, Japan}

\date{}

\maketitle

\begin{abstract}
The symmetric Kullback-Leibler centroid also called the Jeffreys centroid of a set of mutually absolutely continuous probability distributions on a measure space provides a notion of centrality which has proven useful in many tasks including information retrieval, information fusion, and clustering in image, video and  sound processing.
However, the Jeffreys centroid is not available in closed-form for sets of categorical or normal distributions, two widely used statistical models, and thus need to be approximated numerically in practice.
In this paper, we first propose the new Jeffreys-Fisher-Rao center defined as the Fisher-Rao midpoint of the sided Kullback-Leibler centroids as a plug-in replacement of the Jeffreys centroid. This Jeffreys-Fisher-Rao center admits a generic formula for uni-parameter exponential family distributions, and closed-form formula for categorical and normal distributions, matches exactly the Jeffreys centroid for same-mean normal distributions, and is experimentally observed in practice to be close to the Jeffreys centroid. 
Second, we define a new type of inductive centers generalizing the principle of Gauss  arithmetic-geometric double sequence mean for pairs of densities of any given exponential family. This center is shown experimentally to approximate very well the Jeffreys centroid and is suggested to use when the Jeffreys-Fisher-Rao center is not available in closed form.
Moreover, this Gauss-Bregman inductive center always converges and matches  the Jeffreys centroid for sets of same-mean normal distributions.
We report on our experiments demonstrating the use of the Jeffreys-Fisher-Rao and Gauss-Bregman centers instead of the Jeffreys centroid.
Finally, we conclude this work by reinterpreting these fast  proxy centers of Jeffreys centroids under the lens of dually flat spaces in information geometry.
\end{abstract}


\sloppy

\section{Introduction}

Let $(\calX,\calF)$ be a measurable space with sample space $\calX$ and $\sigma$-algebra of events $\calF$, and $\mu$ a positive measure.
We consider a finite set $\{P_1,\ldots,P_n\}$ of $n$  probability distributions all dominated by $\mu$ and  weighted by  a vector $w$ 
belonging to the open standard simplex
$\Delta_n= \left\{ x=(x_1,\ldots,x_n) \st  x_1>0,\ldots,x_n>0, \sum_{i=1}^n x_i=1 \right\}\subset\bbR^n$.
Let $\calP=\{p_1,\ldots,p_n\}$ be the Radon-Nikodym densities of $P_1,\ldots,P_n$ with respect to $\mu$, i.e.,
 $p_i=\frac{\mathrm{d}P_i}{\dmu}$.

The Kullback-Leibler divergence (KLD) between two densities $p(x)$ and $q(x)$ is defined by $D_\KL(p:q)=\int p(x)\log\frac{p(x)}{q(x)}\,\dmu(x)$.
The KLD is asymmetric: $D_\KL(p:q)\not=D_\KL(q:p)$. We use  the argument delimiter  `:' as a notation to indicate this asymmetry.
The Jeffreys divergence~\cite{Jeffreys-1998} symmetrizes the KLD as follows:
\begin{eqnarray*}
D_J(p,q) &=& D_\KL(p:q)+D_\KL(q:p),\\
&=& \int_\calX (p(x)-q(x))\log\frac{p(x)}{q(x)}\dmu(x).
\end{eqnarray*}
 
In general, the $D$-barycenter $C_D$  of $\calP$ with respect to a statistical dissimilarity measure $D(\cdot:\cdot)$  yields a notion of centrality $C_R$ defined by the following optimization problem:
\begin{equation}\label{eq:gencentroid}
c_R = \arg\min_p \sum_{i=1}^n w_i \, D(p_i:p).
\end{equation}
Here, the upper case letter `R' indicates that the optimization defining the $D$-barycenter is carried on the right argument.
When $w=(\frac{1}{n},\ldots,\frac{1}{n})$ is the uniform weight vector, the $D$-barycenter is called the $D$-centroid.
We shall loosely call centroids barycenters in the remainder even when the weight vector is not uniform. 
Centroids with respect to information-theoretic measures have been studied in the literature: 

Let us mention some examples of centroids: 
The entropic centroids~\cite{ben1989entropic} (i.e., Bregman centroids and  $f$-divergences centroids), the Burbea-Rao and Bhattacharyya centroids~\cite{nielsen2011burbea}, the $\alpha$-centroids with respect to $\alpha$-divergences~\cite{Amari-2007}, the Jensen-Shannon centroids~\cite{nielsen2020generalization}, etc.

The $D_J$-centroid is also called the symmetric Kullback-Leibler (SKL) divergence centroid~\cite{Veldhuis-2002} in the literature.
However, since there are many possible symmetrizations of the KLD~\cite{nielsen2019jensen} like the Jensen-Shannon divergence~\cite{lin1991divergence} or the resistor KLD~\cite{johnson2001symmetrizing}, we prefer to use the term Jeffreys centroid instead of SKL centroid to avoid any possible ambiguity on the underlying divergence.
Notice that the square-root of the Jensen-Shannon divergence is a metric distance~\cite{fuglede2004jensen,sra2021metrics} 
but all powers $D_J^\alpha$ of Jeffreys divergence $D_J$ for $\alpha>0$ do not yield metric distances~\cite{vajda2009metric}.

This paper considers the Jeffreys centroids of a finite weighted set of densities $\calP=\{p_{\theta_1},\ldots,p_{\theta_n}\}$ belonging to some prescribed  exponential family~\cite{barndorff2014information} $\calE$:
\begin{equation}\label{eq:Jeffreyscentroid}
c = \arg\min_p \sum_{i=1}^n w_i\, D_J(p_{\theta_i},p).
\end{equation}
In particular, we are interested in computing the Jeffreys centroids for sets of categorical distributions or sets of multivariate normal distributions~\cite{IG-2016}.

\begin{figure}
\begin{tabular}{ll}
\fbox{\includegraphics[width=0.45\textwidth]{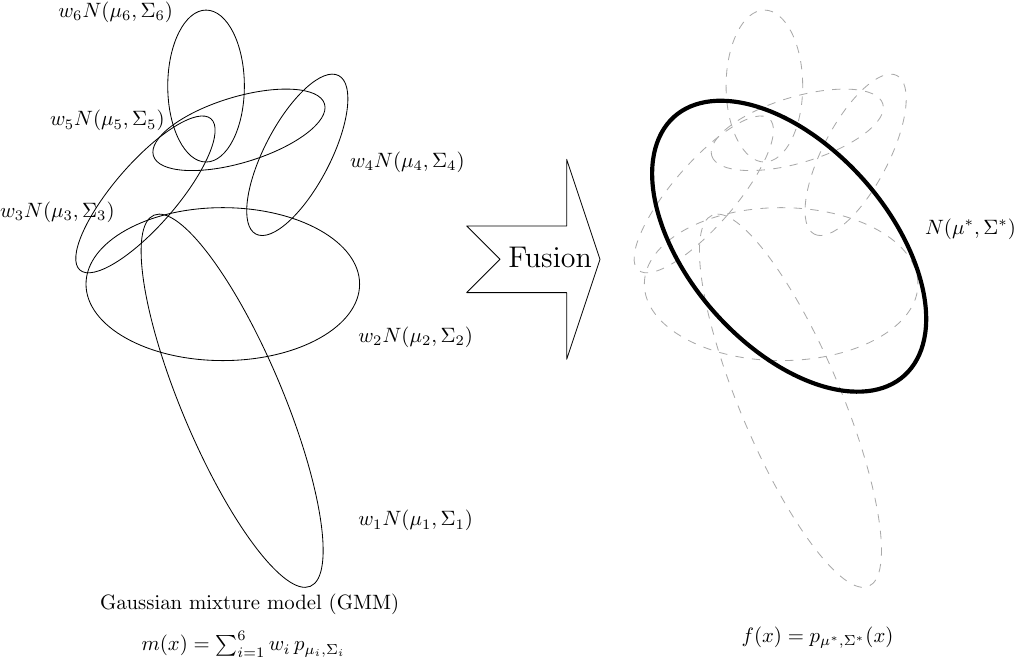}} &
\fbox{\includegraphics[width=0.45\textwidth]{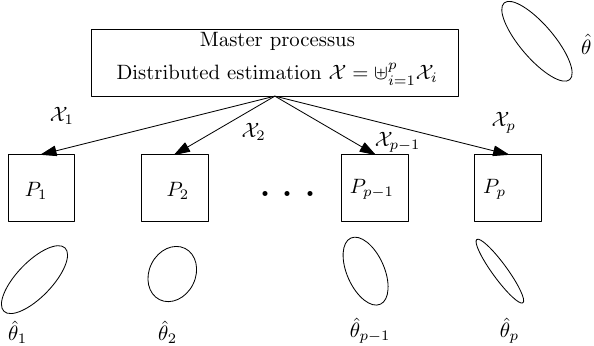}}
\end{tabular}
\caption{Application of centroids and centers in signal processing.
Left: information fusion and mixture model simplification, a Gaussian mixture model is simplified to a single normal distribution.
Right: distributed estimation, a dataset is split among $p$ processus $P_i$'s which first estimate the statistical model parameters $\hat\theta_i$'s. 
Then the processus models are aggregated to yield a single consolidated model $\hat\theta$. }\label{fig:app}
\end{figure}

In general, centroids are used in $k$-means~\cite{lloyd1982least,davis2006differential} type clustering or hierarchical clustering (e.g., Ward criterion~\cite{murtagh2014ward}) and information fusion tasks~\cite{julier2017general} (related to distributed model estimation~\cite{liu2014distributed}) among others.
See Figure~\ref{fig:app}. The choice of the dissimilarity measure depends on the application at hand~\cite{basseville2013divergence}.
Clustering with respect to Jeffreys divergence/Jeffreys centroid has proven useful in many scenario:
For example, it was shown to perform experimentally better than Euclidean or square Euclidean distances for compressed histograms of gradient descriptors~\cite{chandrasekhar2012compressed} or in fuzzy clustering~\cite{seal2020fuzzy}.  
Jeffreys divergence has also been used for image processing~{vasconcelos2004efficient} including image segmentation~\cite{ge2022active}, speech processing~\cite{tabibian2015speech}, computer vision~\cite{zhao2014tensor}, just to name a few.
In particular, finding weighted means of centered normal distributions plays an important role in diffusion tensor imaging (DTI) for smoothing and filtering DT images~\cite{welk2006tensor} which consist of  sets of normal distributions centered at 3D grid locations.

In general, the Jeffreys centroid is not known in closed-form for exponential families~\cite{nielsen2009sided} like the family of categorical distributions or the family of normal distributions often met in applications, and thus need to be numerically approximated in practice.
The main contribution of this paper is to present and study two proxy centers as drop-in replacements of the Jeffreys centroids in applications, and report generic structural formula for generic exponential families with explicit closed-form formula for the families of categorical and normal distributions. 
Namely,  we define the Jeffreys-Fisher-Rao (JFR) center (Definition~\ref{def:JFR}) and the Gauss-Bregman (GB) inductive center 
(Definition~\ref{def:GBcenter}) in Section~\ref{sec:proxy}.

The paper is organized as follows:
By interpreting in two different ways the closed-form formula of the Jeffreys centroids for the particular case of sets of centered normal distributions~\cite{moakher2006symmetric} (proof reported in Appendix~\ref{sec:logdet}), we define the Gauss-Bregman (GB) centers  and the Jeffreys-Fisher-Rao (JFR) centers for sets of densities belonging to an exponential family in Section~\ref{sec:proxy}.
The Jeffreys centroid coincide with both the Gauss-Bregman inductive center and the Jeffreys-Fisher-Rao center for centered normal distributions but differ from each others in general.
In Section~\ref{sec:GBcenter}, we study the Gauss-Bregman inductive center~\cite{sturm2003probability} induced by the cumulant function of an exponential family, and proves the convergence under separability condition of the generalized Gauss double sequences in the limit (Theorem~\ref{thm:convergence}). This Gauss-Bregman center can be easily approximated by   limiting the number of iterations of a double sequence inducing it.
In Section~\ref{sec:JFR}, we report the generic formula for Jeffreys-Fisher-Rao centers for sets of uni-order exponential families, and give explicitly the closed-form formula for the categorical family and the normal family. A comparison of those proxy centers with the numerical Jeffreys centroids is experimentally studied and visually illustrated with some examples.
Thus we propose to use in applications (e.g., clustering) either the fast  Jeffreys-Fisher-Rao center when a closed-form formula is available for the family of distributions at hand or the Gauss-Bregman center approximation with a prescribed number of iterations as a drop-in replacement of the numerical Jeffreys centroids while keeping the Jeffreys divergence.
Some experiments of the JFR and GB centers are reported for the Jeffreys centroid of categorical distributions in Section~\ref{sec:experiments}.
Finally, we conclude the paper in~Section~\ref{sec:concl} with  a discussion  and a generalization of our results to the more general setting of dually flat spaces of information geometry~\cite{IG-2016}.
In Appendix, we give explicitly the algorithm outlined in~\cite{nielsen2013jeffreys} for computing numerically the Jeffreys centroid of sets of categorical distributions in \S\ref{sec:numericalJ}, report a proof on the closed-form formula of the Jeffreys centroid for centered normal distributions~\cite{moakher2006symmetric} in~\S\ref{sec:logdet} that motivated this paper, 
and explain how to calculate in practice the elaborated closed-form formula for the Fisher-Rao geodesic midpoint between two multivariate normal distributions~\cite{kobayashi2023geodesics} in~\S\ref{sec:frmvnmidpoint}.

\section{Proxy centers for Jeffreys centroids}\label{sec:proxy}

\subsection{Background on Jeffreys centroids}

 A density $p_{\theta}$ belonging to an exponential family~\cite{barndorff2014information} $\calE$ can be expressed canonically as $p_\theta(x)=\exp(\inner{\theta}{t(x)}-F(\theta))\, \dmu(x)$, where $t(x)$ is a sufficient statistic vector and $F(\theta)=\log \int \exp(\inner{\theta}{t(x)})\,\dmu(x)$ is the log-normalizer and $\theta$ is the natural parameter belonging to the natural parameter space $\Theta$. We consider minimal regular exponential families~\cite{barndorff2014information} like the discrete family of categorical distributions (i.e., $\mu$ is the counting measure) or the continuous family of multivariate normal distributions (i.e., $\mu$ is the Lebesgue measure).

The Jeffreys centroid of categorical distributions  was first studied by Veldhuis~\cite{Veldhuis-2002} who designed a numerical two-nested loops Newton-like algorithm~\cite{Veldhuis-2002}. 
A random variable $X$ following a categorical distribution $\Cat(p)$ for a parameter $p\in\Delta_d$ in sample space $\calX=\{\omega_1,\ldots, \omega_d\}$ is such that
 $\Pr(X=\omega_i)=p_i$.
Categorical distributions are often used in image processing to statistically model normalized histograms with non-empty bins.
The exact characterization of the Jeffreys centroid was given in~\cite{nielsen2013jeffreys}.
We summarize the categorical Jeffreys centroid in the following theorem~\cite{nielsen2013jeffreys}:

\begin{Theorem}[Categorical Jeffreys centroid~\cite{nielsen2013jeffreys}]\label{thm:sklcat}
The Jeffreys centroid  of a set of $n$ categorical distributions parameterized by $\calP=\{p_1,\ldots,p_n\}\in\Delta_d$ arranged in a matrix $P=[p_{i,j}]\in\bbR^{n\times d}$ and weighted by a vector $w=(w_1,\ldots,w_n)\in\Delta^n$ is $c(\lambda)=(c_1(\lambda),\ldots,c_d(\lambda))$ with
$$
c_j(\lambda)=\frac{a_j}{W_0\left(\frac{a_j}{g_j}\, e^{1+\lambda}\right)}, \quad\forall j\in\{1,\ldots,d\},
$$
where $a_j=\sum_{i=1}^n w_ip_{i,j}$ and $g_j=\frac{\prod_{i=1}^n p_{i,j}^{w_i}}{\sum_{j=1}^d \prod_{i=1}^n p_{i,j}^{w_i} }$ are the $j$-th components of the weighted arithmetic and normalized geometric means, respectively, $W_0$ is the principal branch of the Lambert $W$ function~\cite{corless1996lambert}, and $\lambda\leq 0$ is the unique real value such that $\lambda=-D_\KL(c(\lambda):g)$. 
\end{Theorem}

Furthermore, a simple bisection search is reported in \cite{nielsen2013jeffreys} \S III.B that we convert into Algorithm~\ref{algo:numJ} in the Appendix allows one to approximate numerically the Jeffreys centroid to arbitrary fine precision.

\subsection{Jeffreys centroids on exponential family densities: Symmetrized Bregman centroids}
The Jeffreys divergence between two densities of an exponential family $\calE=\{p_\theta(x)=\exp(\inner{t(x)}{\theta}-F(\theta)) \st\theta\in\Theta\}$ with 
cumulant function $F(\theta)$ amounts to a symmetrized Bregman divergence~\cite{nielsen2009sided} (SBD):
$$
D_J(p_{\theta},p_{\theta'})=S_F(\theta,\theta'):=\inner{\theta_1-\theta_2}{\nabla F(\theta_1)-\nabla F(\theta_2)}.
$$
Using convex duality, we have $S_F(\theta,\theta')=S_{F^*}(\eta,\eta')$, where $\eta=\nabla F(\theta)$ and $F^*(\eta)=\inner{\eta}{(\nabla F)^{-1}(\eta)}-F((\nabla F)^{-1}(\eta))$ is the convex conjugate.
Thus the Jeffreys barycenter of $\calP=\{p_{\theta_1},\ldots,p_{\theta_n}\}$ amounts to either a symmetrized Bregman barycenter on the natural parameters $\calP_\theta=\{\theta_1,\ldots,\theta_n\}$ with respect to $S_F$ or a symmetrized Bregman barycenter on the dual moment parameters $\calP_\eta=\{\eta_1,\ldots,\eta_n\}$ with respect to $S_{F^*}$.
 
It was shown in~\cite{nielsen2009sided} that the symmetrized Bregman barycenter $\theta_S$ of $n$ weighted points amounts to the following minimization problem involving only the sided Bregman centroids:
\begin{eqnarray}
\theta_S &:=& \arg \min_{\theta\in\Theta} \sum_i w_i S_F(\theta,\theta_i),\nonumber\\
&\equiv & \arg\min_{\theta\in\Theta}  B_F(\bartheta:\theta)+B_F(\theta:\ubartheta),\label{eq:reducedBDcentroid}
\end{eqnarray}
where $\bartheta=\sum_i w_i\theta_i$ (right Bregman centroid) and $\ubartheta=(\nabla F)^{-1}(\sum_i w_i\nabla F(\theta_i))$ (left Bregman centroid).
Those $\bartheta$ and $\ubartheta$ centers are centroids~\cite{nielsen2009sided} with respect to the Bregman divergence $B_F(\theta_1:\theta_2)=F(\theta_1)-F(\theta_2)-\inner{\theta_1-\theta_2}{\nabla F(\theta_2)}$ and reverse Bregman divergence: ${B_F}^*(\theta_1:\theta_2):=B_F(\theta_2:\theta_1)$:
\begin{eqnarray*}
\bartheta &=& \arg\min_\theta \sum_i w_i B_F(\theta_i:\theta),\\
\ubartheta &=& \arg\min_\theta \sum_i w_i B_F(\theta:\theta_i)=\arg\min_\theta \sum_i w_i {B_F}^*(\theta_i:\theta).
\end{eqnarray*}

In general, when $H:\bbR^m\rightarrow\bbR$ is a strictly convex differentiable real-valued function of Legendre type~\cite{LegendreType-1967}, the gradient $\nabla H$ is globally invertible (in general the implicit inverse function theorem only guarantees locally the inverse function) and we can define a quasi-arithmetic center of a point set $\calP=\{\theta_1,\ldots,\theta_n\}$ weighted by $w$ as follows:

\begin{Definition}[Quasi-arithmetic center]\label{def:qac}
Let $H=\nabla F$ be the gradient of a  strictly convex or concave differentiable real-valued function $F$ of Legendre type.
The quasi-arithmetic center $c_H(\theta_1,\ldots,\theta_n;w)$ is defined by:
$$
c_H(\theta_1,\ldots,\theta_n;w)=H^{-1}\left(\sum_{i=1}^n w_i H(\theta_i)\right).
$$
\end{Definition}
This definition generalizes the scalar quasi-arithmetic means~\cite{bullen2003quasi} for univariate functions $h$ which are continuous and strictly monotone.
Quasi-arithmetic means are also called $f$-means or Kolmogorov-Nagumo means.
Let $m_{\nabla F}(\theta_1,\theta_2)=c_{\nabla F}(\theta_1,\theta_2;\frac{1}{2},\frac{1}{2})$.

Thus we can solve for $\theta_S$ by setting the gradient of $L(\theta)=B_F(\bartheta:\theta)+B_F(\theta:\ubartheta)$ to zero.
In general, no closed-form formula is known for the symmetrized Bregman centroids and a numerical approximation method was reported in~\cite{nielsen2009sided}.
To circumvent the lack of closed-form formula of symmetrized Bregman centroids for clustering, Nock et al.~\cite{nock2008mixed} proposed a mixed Bregman clustering where each cluster has two representative dual Bregman centroids $\bartheta=\sum_i w_i\theta_i$ (right Bregman centroid) and $\ubartheta=(\nabla F)^{-1}(\sum_i w_i\nabla F(\theta_i))$ (left Bregman centroid), and the dissimilarity measure is a mixed Bregman divergence defined by:
$$
\Delta_F(\theta_1:\theta:\theta_2):=\frac{1}{2}B_F(\theta_1:\theta)+\frac{1}{2}B_F(\theta:\theta_2).
$$ 
Notice that minimizing Eq.~\ref{eq:reducedBDcentroid} amounts to minimize the mixed Bregman divergence:
$$
\min_\theta \Delta_F(\bartheta:\theta:\ubartheta).
$$

However, a remarkable special case is the family of multivariate normal distributions centered at the origin for which the Jeffreys centroid was reported in closed-form 
in~\cite{moakher2006symmetric}.
Let $\calN_0=\{p_{\Sigma}\st \Sigma\in\Sym^{++}(\bbR,d)\}$ be the exponential family with sufficient statistics $t(x)=-\frac{1}{2}(x,xx^\top)$, natural parameter $\theta=\Sigma^{-1}$ (precision matrix) where the covariance matrix belongs to the cone $\Sym^{++}(\bbR,d)$ of symmetric positive-definite matrices, inner product $\inner{X}{Y}=\tr(XY)$, and $F(\theta)=-\frac{1}{2}\log\det(\theta)$. 
In that case, the Jeffreys divergence amounts to a symmetrized Bregman log-det (ld) divergence between the corresponding natural parameters:
$$
D_J(p_\Sigma,p_{\Sigma'})=\frac{1}{2} \, \tr\left(\left({\Sigma'}^{-1}-\Sigma^{-1})(\Sigma-\Sigma'\right)\right)=:\frac{1}{2}S_\ld(\Sigma^{-1},{\Sigma'}^{-1}).
$$
Using the standard covariance matrix parameterization $\Sigma$, we can further express the Jeffreys divergence between two normal distributions 
$p_\Sigma$ and $p_{\Sigma'}$ as:
$$
D_J(p_\Sigma,p_{\Sigma'})=\sum_{i=1}^d \left(\sqrt{\lambda_i}-\frac{1}{\sqrt{\lambda_i}}\right)^2,
$$
where $\lambda_i$'s are the eigenvalues of $\Sigma^{-1}\Sigma'$.
The symmetrized log-det divergence $S_\ld$ is also called the symmetrized Stein loss~\cite{SteinLoss-1961,salehian2013recursive}.
When $d=1$, this divergence is the symmetrized Itakura-Saito divergence also called the COSH distance~\cite{nielsen2009sided}.
The Jeffreys centroid can be characterized using the Fisher-Rao geometry~\cite{Skovgaard-1984} of $\calN_0$ as the Fisher-Rao geodesic midpoint of the sided Kullback-Leibler centroids as follows:

\begin{Theorem}[\cite{moakher2006symmetric}]\label{prop:smmvn}
The Jeffreys centroid $C$ of a set of $n$ centered normal distributions $\calP=\{p_{\Sigma_1},\ldots,p_{\Sigma_n}\}$ weighted with $w_in\Delta_n$ amounts to the symmetrized log-det Bregman centroid for the corresponding weighted set of positive-definite precision matrices  $\calP_\theta=\{P_1=\Sigma_1^{-1},\ldots,P_n=\Sigma_n^{-1}\}$.
The symmetrized log-det Bregman barycenter $C$
 is the Riemannian geodesic midpoint $A\# H$ of the arithmetic barycenter $A=\sum_{i=1}^n w_iP_i$ and harmonic barycenter $H=\left(\sum_{i=1}^n w_i P_i^{-1}\right)^{-1}$
where $X\# Y := X^{\frac{1}{2}}\, \left(X^{-\frac{1}{2}}\, Y\, X^{-\frac{1}{2}} \right)^{\frac{1}{2}}\, X^{\frac{1}{2}}$ is the matrix geometric mean~\cite{bhatia2012riemannian} $G(X,Y)=X\# Y$:
\begin{equation}\label{eq:sldsol}
C= (\sum_{i=1}^n w_iP_i) \# \left(\sum_{i=1}^n w_i P_i^{-1}\right)^{-1}.
\end{equation}
\end{Theorem}

Since the proof of this results involving matrix analysis was omitted in~\cite{moakher2006symmetric}, we report it in full details in Appendix~\ref{sec:logdet}.

Next, we shall define two types of centers for sets of densities of a prescribed exponential family based on two different interpretations of Eq.~\ref{eq:sldsol}.
We call them centers and not centroids because those points are defined by generic structural formula instead of solutions of minimization problems of average divergences of Eq.~\ref{eq:gencentroid}.

\subsection{The Jeffreys-Fisher-Rao center}
Since an exponential family $\calE=\{p_\theta(x)\}$ induces the Riemannian manifold $(\calM,g)$ with Fisher metric $g$ expressed in the $\theta$-parameterization by the Fisher information matrix $\nabla^2 F(\theta)$ and Fisher-Rao geodesics $\gamma(p,q,t)$ defined with respect to the Levi-Civita connection $\bar\nabla$ (induced by $g$), we shall define the Jeffreys-Fisher-Rao center on $\calM$ using the Fisher-Rao geodesics as follows:

\begin{Definition}[Jeffreys-Fisher-Rao (JFR) center]\label{def:JFR}
The Jeffreys-Fisher-Rao center $\theta_\JFR$ of a set $\{p_{\theta_1},\ldots, p_{\theta_n}\}$ of weighted densities by $w\in\Delta_n$ is defined as the Fisher-Rao midpoint of the sided Kullback-Leibler centroids $\bartheta=\sum_i w_i\theta_i$ and $\ubartheta=(\nabla F)^{-1}(\sum_i w_i\nabla F(\theta_i))$: 
\begin{equation}\label{eq:jfrformula}
\theta_\JFR = \bartheta \#\ubartheta,
\end{equation}
where $p\# q=\gamma\left(p,q,\frac{1}{2}\right)$.
\end{Definition}
Equation~\ref{eq:jfrformula} is a generalization of Eq.~\ref{eq:sldsol}: 
Therefore the JFR center matches the Jeffreys centroid for same-mean normal distributions (Theorem~\ref{prop:smmvn}).

\begin{figure}
\begin{tabular}{ll}
\includegraphics[width=0.45\textwidth]{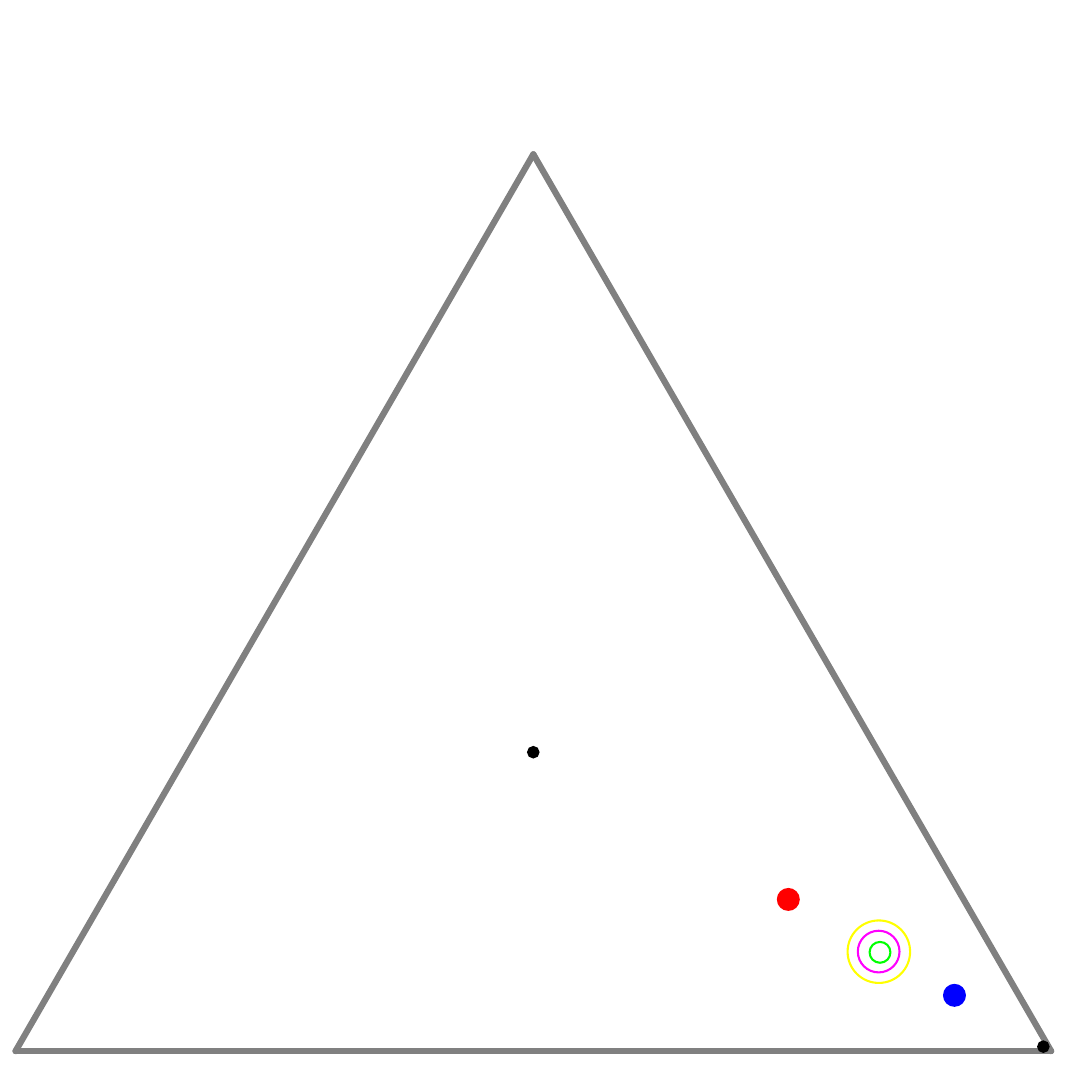} &
\includegraphics[width=0.45\textwidth]{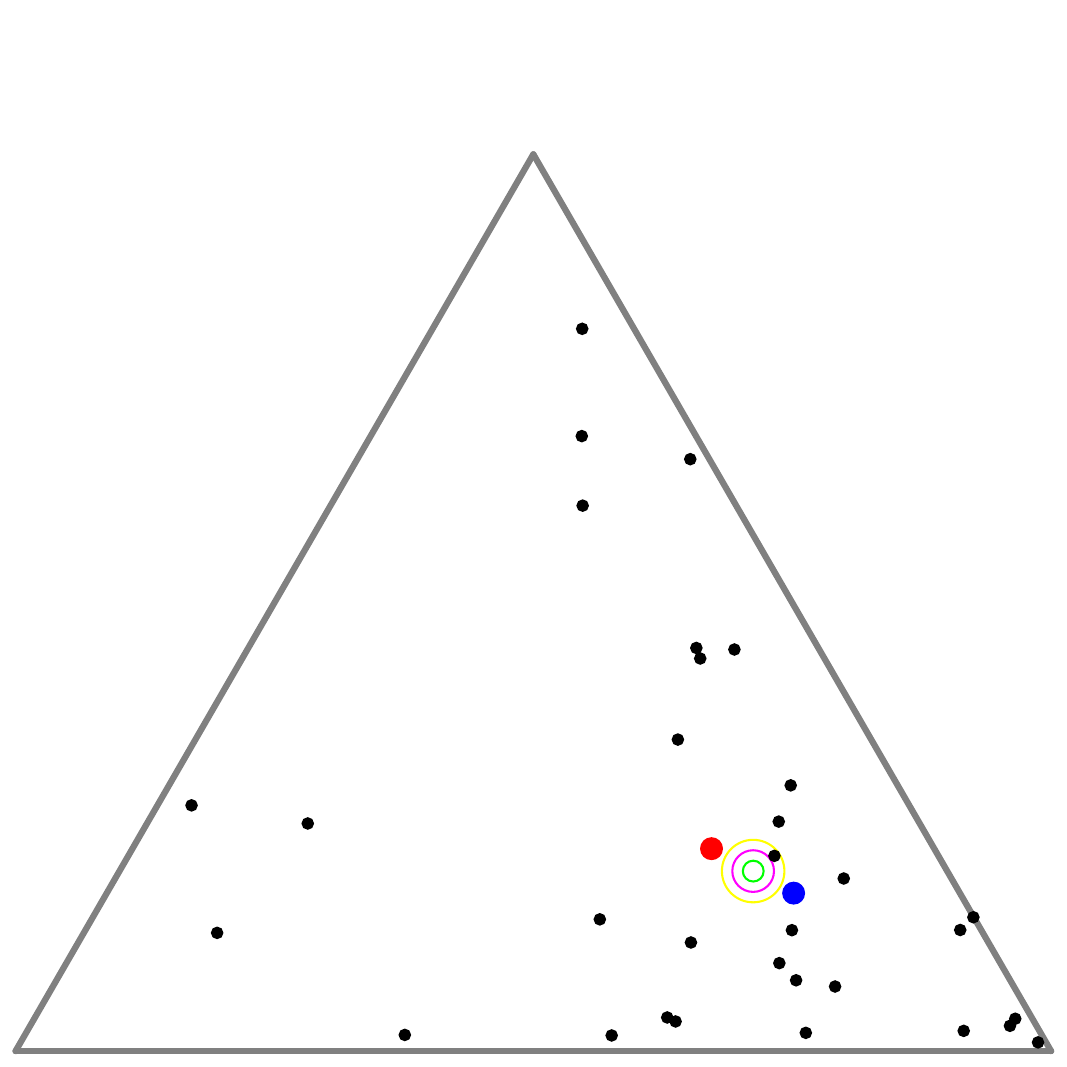} 
\end{tabular}
\caption{Visualizing the arithmetic, normalized geometric, numerical Jeffreys, Jeffreys-Fisher-Rao, and Gauss-Bregman centroids/centers in red, blue, green, purple and yellow, respectively.
Left: Input set consists of $n=32$ trinomial distributions with parameters chosen randomly.
Right: Input set consists of two trinomial distributions with parameters $(\frac{1}{2},\frac{1}{2})$ and $(0.99,0.005,0.005)$.
The numerical Jeffreys centroid (green) is time consuming to calculate using the Lambert $W$ function.
However, the Jeffreys centroid can be well approximated by either the Jeffreys-Fisher-Rao  center (purple) or the inductive Gauss-Bregman center (yellow).
Point centers are visualize with different radii in order to distinguish them easily.
}\label{fig:trinoulli}
\end{figure}

\subsection{Gauss-Bregman inductive center}\label{sec:GBcenter}

Another remarkable property of the Jeffreys  centroid for a set $\{p_{\mu,\Sigma_1},\ldots,p_{\mu,\Sigma_n}\}$ of same-mean normal distributions weighted by $w\in\Delta_n$ with arithmetic and harmonic means $A=\sum_{i=1}^n w_i\Sigma_i^{-1}$ and $H=(\sum_{i=1}^n w_i\Sigma_i)^{-1}$ on the precision matrices $\Sigma_1^{-1},\ldots,\Sigma_n^{-1}$, respectively, is that we have the following invariance of Jeffreys centroid (see Lemma 17.4.4 of \cite{moakher2006symmetric}):
\begin{equation}\label{eq:Ginvariance}
G(A,H)=G\left(\frac{A+H}{2},2\, (A^{-1}+H^{-1})^{-1}\right).
\end{equation}

Nakamura~\cite{nakamura2001algorithms} defined the following double sequence scheme converging to the matrix geometry mean $G(P,Q)$ for any two symmetric 
positive-definite matrices $P$ and $Q$: 
\begin{eqnarray*}
P_{t+1} &=& A(P_t,Q_t) := \frac{P_t+Q_t}{2},\\
Q_{t+1} &=& H(P_t,Q_t) := 2\, (P_t^{-1}+Q_t^{-1})^{-1},
\end{eqnarray*}
initialized with $P_0=P$ and $Q_0=Q$. We have $\lim_{t\rightarrow\infty} P_t=\lim_{t\rightarrow\infty} Q_t= P\# Q=G(P,Q)$.
Let $P_\infty=\lim_{t\rightarrow\infty} P_t$ and $Q_\infty=\lim_{t\rightarrow\infty} Q_t$.
That is, the geometric matrix mean can be obtained as the limits of a double sequence of means.
We can thus approximate $G(P,Q)$ by stopping the double sequence after $T$ iterations to get 
$$
G^{(T)}(P,Q)=A(P_T,Q_T)\approx G(P,Q).
$$

Notice that we can recover those iterations from the invariance property of Eq.~\ref{eq:Ginvariance}:
Indeed, we have 
\begin{equation}\label{eq:reclogdet}
G(P_0,Q_0)=G(\underbrace{A(P_0,Q_0)}_{=:P_1},\underbrace{H(P_0,Q_0)}_{=:Q_1})=G(\underbrace{A(P_1,Q_1)}_{=:P_2},\underbrace{H(P_1,Q_1)}_{=:Q_2})=\ldots,
\end{equation}
and $\|P_t-Q_t\|=\sqrt{\tr((P_t-Q_t)(P_t-Q_t))}$ decreases~\cite{nakamura2001algorithms} at the number of iterations $t$ increases.
Thus by induction $G(P_0,Q_0)=G(P_\infty,Q_\infty)$ with $P_\infty=Q_\infty$. Since $G(X,X)=X$ (means are reflexive), it follows that $G(P_0,Q_0)=P_\infty=Q_\infty$.
It is proved in~\cite{nakamura2001algorithms} that the  convergence rate of the sequence of double iterations is quadratic.
This type of means have been called inductive means~\cite{sturm2003probability,InductiveMean-2023}, and originated from Gauss arithmetic-geometric mean~\cite{almkvist1988gauss}.

Our second interpretation of the geometric matrix mean of Eq.~\ref{eq:sldsol} is to consider it as an inductive mean~\cite{sturm2003probability}, and to generalize this double sequence process to pairs/sets of densities of an exponential family as follows:

\begin{Definition}[Gauss-Bregman $(A,\nabla F)$-center]\label{def:GBcenter}
Let  $\calP=\{p_{\theta_1},\ldots,p_{\theta_n}\}$ be a set of $n$ distributions of an exponential family with cumulant function $F(\theta)$ weighted by a vector $w\in\Delta_n$.
Then the Gauss-Bregman inductive center $\theta_\GB$ is defined as the limit of the double sequence   
\begin{eqnarray*}
\bartheta_{t+1} &=& A(\bartheta_t,\ubartheta_t) := \frac{\bartheta_t+\ubartheta_t}{2},\\
\ubartheta_{t+1} &=& m_{\nabla F}(\bartheta_t,\ubartheta_t) := (\nabla F)^{-1}\left(\frac{\nabla F(\bartheta_t)+\nabla F(\ubartheta_t)}{2}\right),
\end{eqnarray*}
initialized with $\bartheta_{0}=\bartheta=\sum_{i=1}^n w_i\theta_i$ (right Bregman centroid) and 
$\ubartheta_{0}=\bartheta=\nabla F^{-1}\left(\sum_{i=1}^n w_i\nabla F(\theta_i)\right)$ (left Bregman centroid).
That is, we have
\begin{equation}
\theta_\GB=\lim_{t\rightarrow\infty} \bartheta_t=\lim_{t\rightarrow\infty} \ubartheta_t.
\end{equation}
\end{Definition}

Algorithm~\ref{algo:inductiveGB} describes the approximation of the Gauss-Bregman inductive center by stopping the double sequence when the iterated centers are close enough to each others.
We shall prove matching convergence of those $\bartheta_t$ and $\ubartheta_t$ sequences under separability conditions in~\S\ref{sec:GBcenter}.

\begin{algorithm}
\KwIn{A set  $\calP=\{p_{\theta_1},\ldots,p_{\theta_n}\}$ of weighted densities with $w\in\Delta_n$ of an exponential family with cumulant function $F(\theta)$, natural parameters $\theta_i$'s lie in an inner product space $(\Theta,\inner{\cdot}{\cdot})$.} 
\KwIn{The distance is defined as $\|\theta-\theta'\|=\sqrt{\inner{\theta-\theta'}{\theta-\theta'}}$}
\KwIn{A precision parameter $\eps>0$}
\KwOut{A numerical approximation of the symmetrized Bregman centroid}
\tcc{Arithmetic weighted mean on natural parameters}
$\bartheta_0=\sum_{i=1}^n w_i \theta_i$   \;
\tcc{Dual  weighted mean}
$\ubartheta_0=\nabla F^{-1}\left(\sum_{i=1}^n w_i \nabla F(\theta_i)\right)$  \;
$t\leftarrow 0$\;
\tcc{Iterate until close to convergence}
\While{$|\bartheta_t-\ubartheta_t|>\eps$}{
$\bartheta_{t+1}=\frac{\bartheta_t+\ubartheta_t}{2}$   \;
$\ubartheta_{t+1}=\nabla F^{-1}\left(\frac{\nabla F(\bartheta_t)+\nabla F(\ubartheta_t)}{2}\right)$   \;
$t\leftarrow t+1$\;
}
\Return $\bartheta_{t-1}$\;
\caption{Gauss-Bregman inductive center.}
\label{algo:inductiveGB}
\end{algorithm}

For example, the Gauss-Bregman center of two categorical distributions $p=(p_1,\ldots,p_d)$ and $p'=(p_1',\ldots,p_d')$ on a sample space $\calX$ of $d$ elements 
is obtained for the cumulant function $F(\theta)=\log(1+\sum_{i=1}^{d-1} e^{\theta_i})$ with gradient $\nabla F(\theta)=\left[\eta_i=\frac{e^{\theta_i}}{1+\sum_{j=1}^{d-1} e^{\theta_j}}\right]_{i}$ where $\theta=(\theta_1=\log\frac{p_1}{p_d},\ldots,\theta_{d-1}=\log\frac{p_{d-1}}{p_d})$ is the natural parameter.
The reciprocal gradient is $(\nabla F)^{-1}(\eta)=\left[\log\frac{\eta_i}{1-\sum_{j=1}^{d-1}\eta_j}\right]_i$.

We may also compute the Gauss-Bregman center of two categorical distributions $\Cat(p)$ and $\Cat(p')$  using iterations of arithmetic means $a_t$ and geometric normalized means    $g_t$:
\begin{eqnarray*}
a_{t+1}^i &=& A(a_t^i,g_t^i) := \frac{a_t^i+g_t^i}{2},\quad \forall i\in\{1,\ldots,d\}\\
u_{t+1}^i &=&  \sqrt{a_t^ig_t^i}, \quad \forall i\in\{1,\ldots,d\},\\
g_{t+1}^i &=& \frac{u_{t+1}^i}{\sum_{j=1}^d u_{t+1}^j}, \quad \forall i\in\{1,\ldots,d\},
\end{eqnarray*}
where the $u_t$'s are unnormalized geometric means and the $g_t$ are normalized geometric means.
We initialize the sequence with $a_0=p$ and $g_0=p'$, and the Gauss-Bregman center is obtained in the limit: 
$m_\GB^{\mathrm{Cat}}(p,p')=\lim_{t\rightarrow\infty} a_t=\lim_{t\rightarrow\infty} g_t$.
See Algorithm~\ref{algo:inductiveGBcat}.

The Jeffreys centroid of a set of centered normal distributions is the Gauss-Bregman center obtained for the generator $F(\theta)=-\frac{1}{2}\log\det(\theta)$, the cumulant function of the exponential family of centered normal distributions.

\begin{algorithm}
\KwIn{A set of weighted categorical distributions: $\calP^w=\{p_1,\ldots,p_n\}$ with $w\in\Delta_n$ and $p_i\in\Delta_d$. 
Let $p_{i,j}$ denote the $j$-th component of $p_i$.}
\KwIn{A precision parameter $\eps>0$}
\KwIn{Distance is chosen as total variation $\frac{1}{2}\|\cdot\|_1$}
\KwOut{A numerical approximation of the SKL centroid/Jeffreys centroid $c$}
\tcc{Arithmetic weighted mean (normalized)}
$a^j_{0}=\sum_{i=1}^n w_ip_{i,j}$ for $i\in\{1,\ldots,d\}$ for $j\in\{1,\ldots,d\}$ \;
\tcc{Normalized geometric weighted mean}
$g^j_{0}=\frac{\prod_{i=1}^n p_{i,j}^{w_i}}{\sum_{j=1}^d \prod_{i=1}^n p_{i,j}^{w_i} }$ for $j\in\{1,\ldots,d\}$ \;
$t\leftarrow 0$\;
\tcc{Iterate until close to convergence}
\While{$\|a_t-g_t\|_1 > 2\eps$}{
\tcc{Arithmetic mean}
$a_{t+1}^i = \frac{a_t^i+g_t^i}{2},\quad \forall i\in\{1,\ldots,d\}$\\
\tcc{Non-normalized geometric mean}
$u_{t+1}^i =  \sqrt{a_t^ig_t^i}, \quad \forall i\in\{1,\ldots,d\}$\\
\tcc{Normalized geometric mean}
$g_{t+1}^i = \frac{u_{t+1}^i}{\sum_{j=1}^d u_{t+1}^j}, \quad \forall i\in\{1,\ldots,d\}$\\
$t\leftarrow t+1$\;
}
\Return $a^{(t-1)}$\;
\caption{Gauss-Bregman inductive center for sets of categorical distributions.}
\label{algo:inductiveGBcat}
\end{algorithm}

Figure~\ref{fig:trinoulli} displays the arithmetic, normalized geometric, numerical Jeffreys, Jeffreys-Fisher-Rao, and Gauss-Bregman centroids/centers
 for a set of $32$ trinomial distributions.
We may consider normalized intensity histograms of images (modeled as multinomials with one trial) quantized with $d=256$ bins: That is, a normalized histogram with $d$ bins is interpreted as a point in $\Delta_d$ and visualized as a polyline with $d-1$ line segments.
Figure~\ref{fig:histogram} (left) displays the various centroids and centers obtained for an input set consisting of two histograms (the commonly use {\tt Barbara} and {\tt Lena} images which have been used in~\cite{nielsen2013jeffreys}).
Notice that the JFR center (purple) and GB center (yellow) are closed to the numerical Jeffreys centroid (green). 
We also provides a close-up window in Figure~\ref{fig:histogram} (right). 

\def\xxx{0.343}
\begin{figure}
\centering
\begin{tabular}{cc}
\includegraphics[height=\xxx\textwidth]{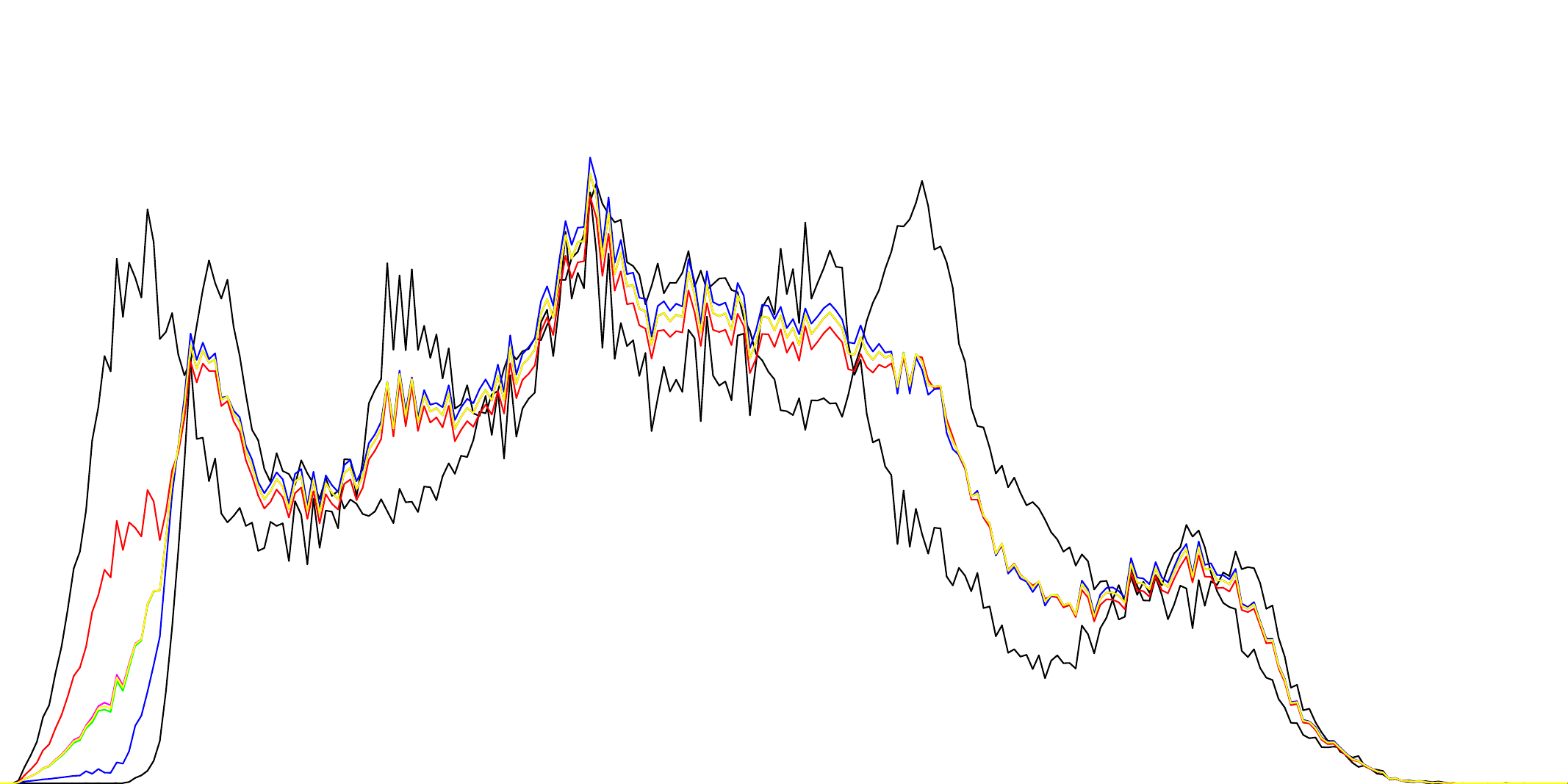} &
\fbox{\includegraphics[height=\xxx\textwidth]{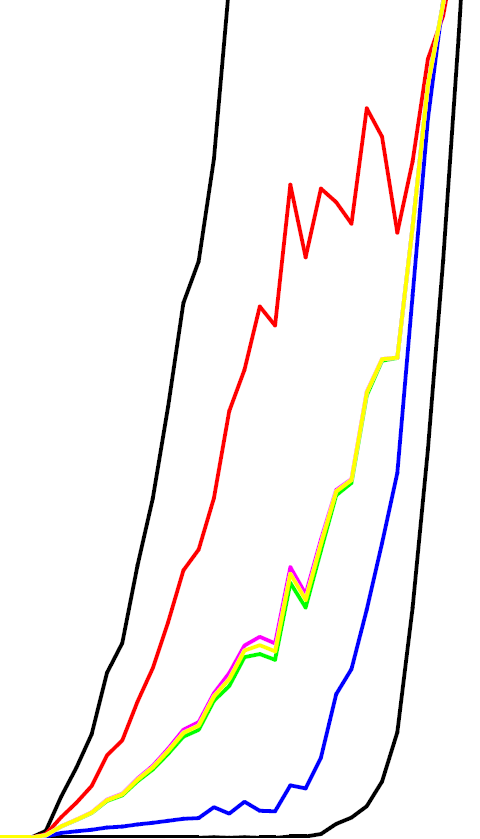}}
\end{tabular}
\caption{Left: Displaying the arithmetic, normalized geometric, numerical Jeffreys, Jeffreys-Fisher-Rao, and Gauss-Bregman centroids/centers in red, blue, green, purple and yellow, respectively. Input set are two normalized histograms with $d=256$ bins plotted as polylines with $255$ line segments.
Observe that the Jeffreys-Fisher-Rao center (purple) and Gauss-Bregman center (yellow) approximates well the Jeffreys centroid (green) which is more computationally expensive to calculate. Right: Closed-up window on the first left bins of normalized histograms.}\label{fig:histogram} 
\end{figure}

Notice that we can check experimentally the quality of the approximation of the Gauss-Bregman center to the Jeffreys centroid, by defining the symmetrized Bregman centroid energy:
$$
E_F(\theta):=\inner{\theta-\bartheta}{\nabla F(\theta)}-\inner{\theta}{\nabla F(\bartheta)},
$$
and checking that $\nabla E_F(\theta)$:
\begin{eqnarray}
\forall i,&& \partial_i \left( \sum_{i=1}^d (\theta_i-\bartheta_i)\partial_i F(\theta) - \theta_i\partial_i F(\bartheta) \right)=0,\\
 && \partial_i F(\theta)+(\theta_i-\bartheta_i) \partial_i^2 F(\theta)-\partial_i F(\bartheta)
+\left( \sum_{j\not =i} (\theta_j-\bartheta_j)\partial_i\partial_j F(\theta) - \partial_i\theta_j\partial_j F(\bartheta) \right)=0
\end{eqnarray}
is close to zero, where $\partial_l:=\frac{\partial}{\partial\theta_l}$.

Next, we study these two new types of centers and how well they approximate the Jeffreys centroid.

\section{Gauss-Bregman inductive centers: Convergence analysis and properties}\label{sec:GBcenteranalysis}

Let $F(\theta)$ be a strictly convex and differentiable real-valued function of Legendre type~\cite{banerjee2005clustering} define on an open parameter space $\Theta$.
Then the gradient map $\theta\mapsto \eta(\theta)=\nabla F(\theta)$ is a bijection with reciprocal inverse function ]
$\eta\mapsto \theta(\eta)=\nabla F^*(\eta)=(\nabla F)^{-1}(\eta)$ where 
$F^*(\eta)=\inner{\eta}{\nabla F^{-1}(\eta)} -F(\nabla F^{-1}(\eta))$ is the Legendre-Fenchel convex conjugate.
For example, we may consider the cumulant functions of regular exponential families.

\begin{figure}
\centering
\includegraphics[width=0.65\textwidth]{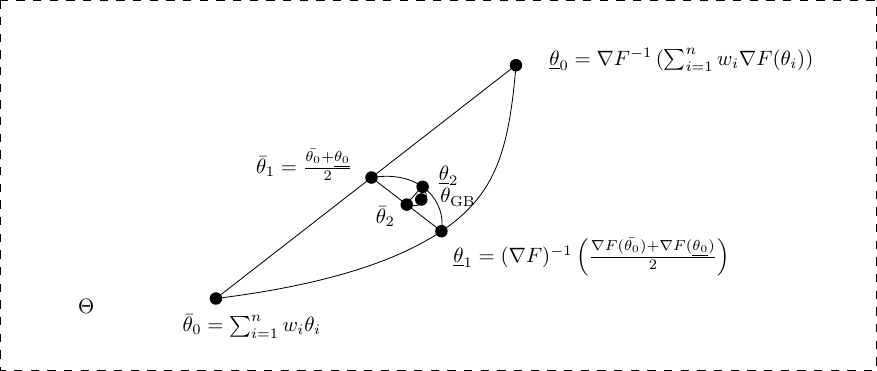}
\caption{Geometric illustration of the double sequence inducing a Gauss-Bregman center in the limit.}\label{fig:GBmean}
\end{figure}

We define the Gauss-Bregman center $\theta_\GB$ of a set $\{\theta_1,\ldots,\theta_n\}$ weighted by $w\in\Delta_n$ as the limits of the sequences $\bartheta_1,\ldots$ and $\ubartheta_1,\ldots$ defined by
\begin{eqnarray}
\bartheta_{t+1} &=& A(\bartheta_t,\ubartheta_t) := \frac{\bartheta_t+\ubartheta_t}{2},\\ \label{eq:iter1}
\ubartheta_{t+1} &=& m_{\nabla F}(\bartheta_t,\ubartheta_t) := (\nabla F)^{-1}\left(\frac{\nabla F(\bartheta_t)+\nabla F(\ubartheta_t)}{2}\right),\label{eq:iter2}
\end{eqnarray}
initialized with $\bartheta_{0}=\bartheta=\sum_{i=1}^n w_i\theta_i$ and $\ubartheta_{0}=\bartheta=\nabla F^{-1}\left(\sum_{i=1}^n w_i\nabla F(\theta_i)\right)$.
That is, we have
$$
\theta_\GB=\lim_{n\rightarrow\infty} \bartheta_{t}=\lim_{n\rightarrow\infty} \ubartheta_{t}.
$$
Such a center has been called an inductive mean by Sturm~\cite{sturm2003probability}. See~\cite{InductiveMean-2023} for an overview of inductive means.
Figure~\ref{fig:GBmean} illustrates geometrically the double sequence iterations converging to the Gauss-Bregman mean.

\begin{Theorem}\label{thm:convergence}
The Gauss-Bregman $(A,\nabla F)$-center with respect to a Legendre type function $F(\theta)$ is well-defined (i.e., double sequence converges) for separable Bregman generators.
\end{Theorem}

\begin{figure}
\centering
\includegraphics[width=0.4\textwidth]{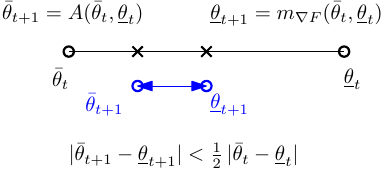}
\caption{Illustration of the double sequence convergence for scalar Gauss-Bregman $(A,m_{\nabla F})$-mean.}\label{fig:GBmean1dconv}
\end{figure}

\begin{Proof}
We need to prove the convergence of $\{\bartheta_t\}$ and $\{\ubartheta_t\}$ to the same finite limit.
When $F(\theta)$ is univariate, the convergence of the inductive centers has been reported in~\cite{lehmer1971compounding}.
We need to prove that the double iterations of Eq.~\ref{eq:iter1} and Eq.~\ref{eq:iter2} converges.

Let us consider the following cases:
\begin{enumerate}

\item When the dimension is one, the quasi-arithmetic mean $m_{f'}$ for $f$ a strictly convex and differentiable function lies between the minimal and maximal argument (i.e., this is the definition of a strict mean):
$$
\min\{\theta_1,\theta_2\}\leq m_{f'}(\theta_1,\theta_2)\leq \max\{\theta_1,\theta_2\}.
$$
Thus we have 
$$
|\bartheta_{t+1}-\ubartheta_{t+1}|\leq \frac{1}{2}|\bartheta_{t}-\ubartheta_{t}|,
$$ 
and it follows that $|\bartheta_{t+1}-\ubartheta_{t+1}|\leq \frac{1}{2^t} |\bartheta_{0}-\ubartheta_{0}|$.
Thus we have quadratic convergence of scalar $(A,f')$-means. See Figure~\ref{fig:GBmean1dconv}.

\item When $F(\theta)$ is multivariate and separable, i.e., $F(\theta)=\sum_{i=1}^d f_i(\theta^i)$ where $\theta=(\theta^1,\ldots,\theta^d)$ are the components of $\theta\in\bbR^d$ and the $f_i$'s are scalar strictly convex and differentiable functions, we can apply  case~1 dimension-wise to get the quadratic convergence. 
 
\item Otherwise, we consider the multivariate quasi-arithmetic center $m_{\nabla F}(\theta,\theta')$ with uniform weight vector $w=(\frac{1}{2},\frac{1}{2})$.
One problem we face is that the quasi-arithmetic center $m_{\nabla F}(\theta,\theta')$ for $\theta\not=\theta'$ may lie outside the open bounding box of $\bbR^d$ with diagonal corners $\theta$ and $\theta'$  
$$
\theta_m=(\min\{\theta^1,{\theta'}^1\},\ldots, \min\{\theta^d,{\theta'}^d\}),\quad \theta_M=(\max\{\theta^1,{\theta'}^1\},\ldots, \max\{\theta^d,{\theta'}^d\}).
$$

Indeed, in the 2D case, we may consider $\theta=(x,y)$ and $\theta'=(x',y)$. Clearly, the open bounding box is empty, and the midpoint $m_{\nabla F}(\theta,\theta')$ lies outside this box. Yet, we are interested in the convergence rate when $\theta'\approx\theta$.

In general, we shall measure the difference between two iterations by the squared norm distance induced by the inner product:
$$
\|A(\theta,\theta')-m_{\nabla F}(\theta,\theta')\|^2 = \inner{A(\theta,\theta')-m_{\nabla F}(\theta,\theta')}{A(\theta,\theta')-m_{\nabla F}(\theta,\theta')}.
$$

\end{enumerate}
\end{Proof}

Let $m^\GB_F(\theta_1,\theta_2)$ denote the Gauss-Bregman center of $\theta_1$ and $\theta_2$, $A(\theta_1,\theta_2)=\frac{\theta_1+\theta_2}{2}$
 the arithmetic mean and $m_{\nabla F}(\theta_1,\theta_2)=(\nabla F)^{-1}\left(\frac{\nabla F(\theta_1)+\nabla F(\theta_2)}{2}\right)$ the quasi-arithmetic center.

By construction, the Gauss-Bregman center enjoys the following invariance property generalizing   Lemma 17.4.4 of \cite{moakher2006symmetric} in the case of the log det generator:

\begin{Property}\label{prop:GBinvariance}
We have $m^\GB_F(\theta_1,\theta_2)=m^\GB_F\left(A(\theta_1,\theta_2),m_{\nabla F}(\theta_1,\theta_2)\right)$.
\end{Property}

\begin{Proof}
Similar to the cascaded inequalities of Eq.~\ref{eq:reclogdet}, we have
\begin{equation} 
m^\GB_F(\theta_1,\theta_2)=m_\GB^F(\underbrace{A(\theta_1,\theta_2)}_{=:\theta_1^{(1)}},\underbrace{m_{\nabla F}(\theta_1,\theta_2)}_{=\theta_2^{(1)}})=\ldots
\end{equation}

In the limit $t\rightarrow\infty$, 
we have $m^\GB_F(\theta_1,\theta_2)=m^\GB_F(\theta_1^{(\infty)},\theta_2^{(\infty)})=m^\GB_F(\theta_1^{(\infty-1)},\theta_2^{(\infty-1)})=\dots$. Since $\infty-1=\infty$, we get the desired invariance property: 
$$m^\GB_F(\theta_1,\theta_2)=m^\GB_F\left(A(\theta_1,\theta_2),m_{\nabla F}(\theta_1,\theta_2)\right).$$
\end{Proof}
 
Note that when $F(\theta)$ is univariate, the Gauss-Bregman mean $m^\GB_F(\theta_1,\theta_2)$ converges at quadratic rate~\cite{lehmer1971compounding}.
In particular, when $F(\theta)=-\log\theta$ (Burg negentropy), we have $F'(\theta)=-\frac{1}{\theta}$ ($m_{F'}$ is the harmonic mean) and the  Gauss-Bregman mean is the arithmetic-harmonic mean (AHM) which converges to the geometric mean, a simple closed-form formula.
Notice that the geometric mean $g=\sqrt{xy}$ of two scalars $x>0$ and $y>0$ can  be expressed using the arithmetic mean $a=\frac{x+y}{2}$ and the harmonic mean $h=\frac{2xy}{x+y}$: $g=\sqrt{ah}$.
But when $F(\theta)=\theta\log\theta-\theta$ (Shannon negentropy), the Gauss-Bregman mean $m^\GB_F(\theta_1,\theta_2)$ coincides with Gauss arithmetic-geometric mean~\cite{almkvist1988gauss}   (AGM) since 
$F'(\theta)=\log\theta$ and $m_{F'}(\theta_1,\theta_2)=\sqrt{\theta_1\theta_2}$, the geometric mean.
Thus $m^\GB_F(\theta_1,\theta_2)$ is related to the elliptic integral $K$ of the first type~\cite{almkvist1988gauss}:
There is no-closed form formula for the AGM in terms of elementary functions as this induced mean  
 is related to the complete elliptic integral of the first kind $K(\cdot)$:
\begin{equation}\label{eq:agm}
\AGM(x,y)=\frac{\pi}{4} \frac{x+y}{K\left(\frac{x-y}{x+y}\right)},
\end{equation}
where $K(u)=\int_0^{\frac{\pi}{2}} \frac{\mathrm{d}\theta}{\sqrt{1-u^2\sin^2(\theta)}}$ is the elliptic integral.
Thus it is difficult, in general, to report closed-form formula for the inductive Gauss-Bregman means even for univariate generators $F(\theta)$.

The Jeffreys centroid of $x>0$ and $y>0$ with respect to the scalar Jeffreys divergence $D_J(p,q)=(p-q)\log\frac{p}{q}$ admits a closed-form solution~\cite{nielsen2013jeffreys}:
\begin{equation}\label{eq:j1dcat}
c=\frac{a}{W_0\left(\frac{a}{g}e\right)}
\end{equation}
where $a=\frac{x+y}{2}$ and $g=\sqrt{xy}$, and $W_0$ is the principal branch of the Lambert $W$ function~\cite{corless1996lambert}. This examples shows that the Gauss-Bregman center does not coincide with Jeffreys centroid in general (e.g., compare Eq.~\ref{eq:agm} with Eq.~\ref{eq:j1dcat}).

\section{Jeffreys-Fisher-Rao centers: Generic structural formula and some closed-form formula}\label{sec:JFR}

\subsection{Jeffreys-Fisher-Rao center for uni-parametric statistical models}\label{sec:JFRuni}

Consider a set $\calP=\{p_{\theta_1},\ldots,p_{\theta_n}\}$ of $n$ parametric distributions  where $\theta\in\Theta\subset\bbR$ is scalar parameter.
Let $w=(w_1,\ldots,w_n)\in\Delta_n$ be a weight vector on $\calP$ such that the weight of $p_{\theta_i}$ is $w_i$.
The distributions $p_{\theta}$'s may not necessarily belong to an exponential family (e.g., Cauchy scale family).
The Fisher-Rao geometry~\cite{miyamoto2024closed,NIELSEN2024} of the parametric family of distributions $\calF=\{p_\theta\st\theta\in\Theta\}$ (statistical model) can be modeled as a Riemannian manifold with Fisher metric 
$g(\theta)=I(\theta)$ defined by the Fisher information $I(\theta)=E_\theta[(\log p_\theta(x))^2]=-E_\theta[\nabla^2 \log p_\theta(x)]$. 
When $\calF$ is an exponential family with cumulant function $F(\theta)$, we have $I(\theta)=F''(\theta)$.

The underlying geometry of $(\calF,g(\theta)=I(\theta))$ is Euclidean after a change of variable $\eta(\theta)=\sqrt{I(\theta)}$ since we can write the metric tensor as follows:
$$
g(\theta)=\sqrt{I(\theta)} \,\times {\underbrace{1}_{= g_{\mathrm{Euclidean}}} },\times\, \sqrt{I(\theta)}.
$$
Thus the Riemannian Fisher-Rao distance is the Euclidean distance expressed in the $h(\theta)$-coordinate system with
 $h(\theta)=\int_{\theta_0}^\theta \sqrt{I(u)}\, \mathrm{d}u$, and we have the Fisher-Rao distance given by:
$$
\rho(p_{\theta_1},p_{\theta_2})=|h(\theta_1)-h(\theta_2)|.
$$

When $\calF$ is an exponential family with cumulant function $f(\theta)$, we have $I(u)=f''(u)$.
 
We summarize the result on the JFR center in the following theorem:

\begin{Theorem}[Jeffreys-Fisher-Rao centroid in uni-order exponential families]\label{thm:Jef1d}
The Jeffreys-Fisher-Rao centroid $\theta_S$ of $n$ densities $p_{\theta_1},\ldots,p_{\theta_n}$ of an exponential family of order $1$  with log-normalizer $f(\theta)$ for $\theta\in\Theta$ the natural parameter space,  and weight vector $w\in\Delta_n$ is
\begin{equation}
\theta_S=m_h(\bartheta,\ubartheta),
\end{equation}
where $m_h(\bartheta,\ubartheta)=h^{-1}\left(\frac{h(\bartheta)+h(\ubartheta)}{2}\right)$ is the quasi-arithmetic mean~\cite{bullen2003quasi} of 
the dual left and right KL centroids $\bartheta=\sum_{i=1}^n w_i\theta_i=\theta_R$ and $\ubartheta=(f')^{-1}(\sum_{i=1}^n w_i f'(\theta_i))$ with respect to the scalar monotone function $h=\int^\theta_{\theta_0}\sqrt{f''(u)}\,\du$ for any $\theta\in\Theta$.
\end{Theorem}

\begin{Proof}
Since the Fisher information is $I(\theta)=f''(\theta)$, we have $h(\theta)=\int_{\theta_0}^\theta \sqrt{f''(u)}\mathrm{d}u$.
The Riemannian center of mass~\cite{karcher1977riemannian} minimizes
$$
\theta_S=\arg\min_\theta \sum_{i=1}^n w_i \rho^2(\theta_i,\theta).
$$

But in the $h$-parameterization,  the Riemannian centroid amounts to an Euclidean center of mass/centroid in the $h$-Cartesian coordinate system:
$$
h(\theta_S) = \sum_{i=1}^n w_i h(\theta_i).
$$

Therefore, we have $\theta_S=h^{-1}(\sum_i w_i h(\theta_i))=:m_h(\theta_1,\ldots,\theta_n;w_1,\ldots,w_n)$, a weighted quasi-arithmetic mean.
Since the Jeffreys centroid amounts to a symmetrized Bregman centroid of the left and right Bregman centroids~\cite{nielsen2009sided}, 
$\ubartheta=m_{f'}(\theta_1,\ldots,\theta_n;w_1,\ldots,w_n)$ and $\bartheta=\sum_i w_i\theta_i$.
It follows that the Jeffreys-Fisher-Rao center is $\theta_\JFR=m_h(\bartheta,\ubartheta)$ after using Property~\ref{eq:reducedBDcentroid}.
\end{Proof}

\subsection{Jeffreys-Fisher-Rao center for categorical distributions}\label{sec:JFRcat}
Recall from Theorem~\ref{thm:sklcat} that the Jeffreys centroid $c=(c_1,\ldots,c_j,\ldots,c_d)$ of a set of $n$ categorical distributions with parameters arranged in the matrix $[p_{i,j}]$ is given by
$$
c_j(\lambda)=\frac{a_j}{W_0\left(\frac{a_j}{g_j}\, e^{1+\lambda}\right)}, \quad\forall j\in\{1,\ldots,d\},
$$
where $a_j=\sum_{i=1}^n w_ip_{i,j}$ and $g_j=\frac{\prod_{i=1}^n p_{i,j}^{w_i}}{\sum_{j=1}^d \prod_{i=1}^n p_{i,j}^{w_i} }$ are the components of the weighted arithmetic and normalized geometric means, respectively and $W_0$ is the principal branch of the Lambert $W$ function~\cite{corless1996lambert}.
The optimal $\lambda\leq 0$ is unique and satisfies $\lambda=-D_\KL(c_j(\lambda):g)$.

Let $c(\lambda)=(c_1(\lambda),\ldots,c_d(\lambda))$.
Let $L_J(p)$ denotes the Jeffreys loss function to minimize to find the optimal Jeffreys centroid:
\begin{equation}\label{eq:infoJ}
L_J(p)=\sum_{i=1}^n w_i D_J(p_i,p)
\end{equation}

We say that $p$ is a $(1+\eps)$-approximation of the exact Jeffreys centroid $c$ when we have
$$
L_J(c)\leq L_J(p)\leq (1+\eps) L_J(c).
$$
It was shown in~\cite{nielsen2013jeffreys} that $\tilde{c}=c(0)$ called the unnormalized Jeffreys center yields a $s(\lambda)-1$-approximation on $c$ where 
$s(\lambda)=\sum_j c_j(\lambda)\leq 1$.

Since the Fisher-Rao geodesic midpoints on the categorical Fisher-Rao manifold are known in closed-form~\cite{vcencov1978algebraic}, we give the mathematical expression of JFR center as follows:

\begin{Theorem}[JFR centroid of categorical distributions]\label{thm:Jcat}
Let $\calP_w=\{p_1,\ldots,p_n\}$ be a set of $n$  probability mass functions weighted by $w\in\Delta_n$ with $p_i=(p_{i,1},\ldots,p_{i,d})\in\Delta_d$ for $i\in\{1,\ldots,n\}$ and $w\in\Delta_n$.
Then the JFR barycenter $c$ minimizing is unique and given by the following formula:

\begin{equation}\label{eq:J}
c_j=\frac{(\sqrt{a_j}+\sqrt{g_j})^2}{2\,(1+\sum_{l=1}^d \sqrt{a_j}\sqrt{g_j})}, \forall j\in\{1,\ldots, d\},
\end{equation}
where $a=(a_1,\ldots,a_d)=\sum_{i=1}^n w_ip_i$ is the weighted arithmetic mean and 
$g=(g_1,\ldots,g_d)$ is the normalized weighted geometric mean with components 
$g_j=\frac{\prod_{i=1}^n p_{i,j}^{w_i}}{\sum_{j=1}^d \prod_{i=1}^n p_{i,j}^{w_i} }$ for $i\in\{1,\ldots,d\}$.
\end{Theorem}

Notice that the JFR center differs from the Jeffreys centroid which requires the use of Lambert $W$ function~\cite{corless1996lambert}. 
However, we noticed that for practical applications, the JFR centroid approximates well the Jeffreys centroid and is much faster to compute (see experiments in Section~\ref{sec:experiments}).

\subsection{Jeffreys-Fisher-Rao center for multivariate normal distributions}\label{sec:JFRmvn}

Let $\calP=\{ p_{\mu_1,\Sigma_1},\ldots, p_{\mu_n,\Sigma_n} \}$ be a set of $n$ probability density functions (PDFs) of $d$-variate normal distributions weighted by $w\in\Delta_n$ where the PDF of a multivariate normal distribution of mean $\mu$ and covariance matrix $\Sigma$ is given by:
$$
{p}_{\mu,\Sigma}=\frac{1}{(2\pi)^{\frac{d}{2}}\sqrt{\det(\Sigma)}} \, \exp\left(-\frac{1}{2}(x-\mu)^\top\Sigma^{-1}(x-\mu)\right).
$$ 

Let $\lambda_i=(\mu_i,\Sigma_i)$ be the ordinary parameterization of normal distributions $p_{\mu_i,\Sigma_i}$.
The family $\calF=\{p_{\mu,\Sigma}(x) \st \mu\in\bbR^d,\Sigma\in\Sym^{++}(\bbR,d)\}$ of normal distributions forms an exponential family with the dual natural $\theta$- and moment $\eta$-parameterizations~\cite{nielsen2019jensen} given by :
\begin{eqnarray*}
\theta(\lambda) &=&  (\theta_v,\theta_M )=\left(\Sigma^{-1}\mu,\frac{1}{2}\Sigma^{-1}\right),\\
\eta(\lambda) &=& \left(\mu,\mu\mu^\top+\Sigma\right),
\end{eqnarray*}
when choosing the sufficient statistic $t(x)=(x,xx^\top)$.
The Jeffreys divergence between two $d$-variate normal distributions $N(\mu_1,\Sigma_1)$ and $N(\mu_2,\Sigma_2)$ is given by the formula
$$
D_J({p}_{\mu_1,\Sigma_1},{p}_{\mu_2,\Sigma_2}) = 
(\mu_2-\mu_1)\top (\Sigma_1^{-1}+\Sigma_2^{-1})(\mu_2-\mu_1) + \tr\left(\Sigma_1^{-1}\Sigma_2+\Sigma_2^{-1}\Sigma_1\right)-2d.
$$

The left and right Kullback-Leibler barycenters  amount to corresponding right and left Bregman barycenters~\cite{nielsen2009sided} induced by the cumulant function
$$
F(\theta)=F(\theta_v,\theta_M)=\frac{1}{2}\left(d\log\pi -\log\det(\theta_M)+\frac{1}{2}\theta_v^\top\theta_M^{-1}\theta_v\right),
$$
and the gradient of $F(\theta)$ defines the dual moment parameter with
$$
\eta(\theta)=\nabla F(\theta)=\left(\frac{1}{2}\theta_M^{-1}\theta_v,\frac{1}{2}\theta_M^{-1}-\frac{1}{4}(\theta_M^{-1}\theta_v)(\theta_M^{-1}\theta_v)^\top\right).
$$
The reciprocal gradient is given by
$$
\theta(\eta)=\theta(\eta_v,\eta_M)=(\nabla F)^{-1}(\eta)=\left(\theta_v=-(\eta_M+\eta_v\eta_v^\top)^{-1}\eta_v,\theta_M=-\frac{1}{2}(\eta_M+\eta_v\eta_v^\top)^{-1}\right).
$$

The Gauss-Bregman center is a $(A,m_{\nabla F})$-inductive center which can be approximated by carrying a prescribed number $T$ of iterations of the Gauss-Bregman double sequence.

\begin{figure}
\centering
\begin{tabular}{ll}
\fbox{\includegraphics[width=0.35\textwidth]{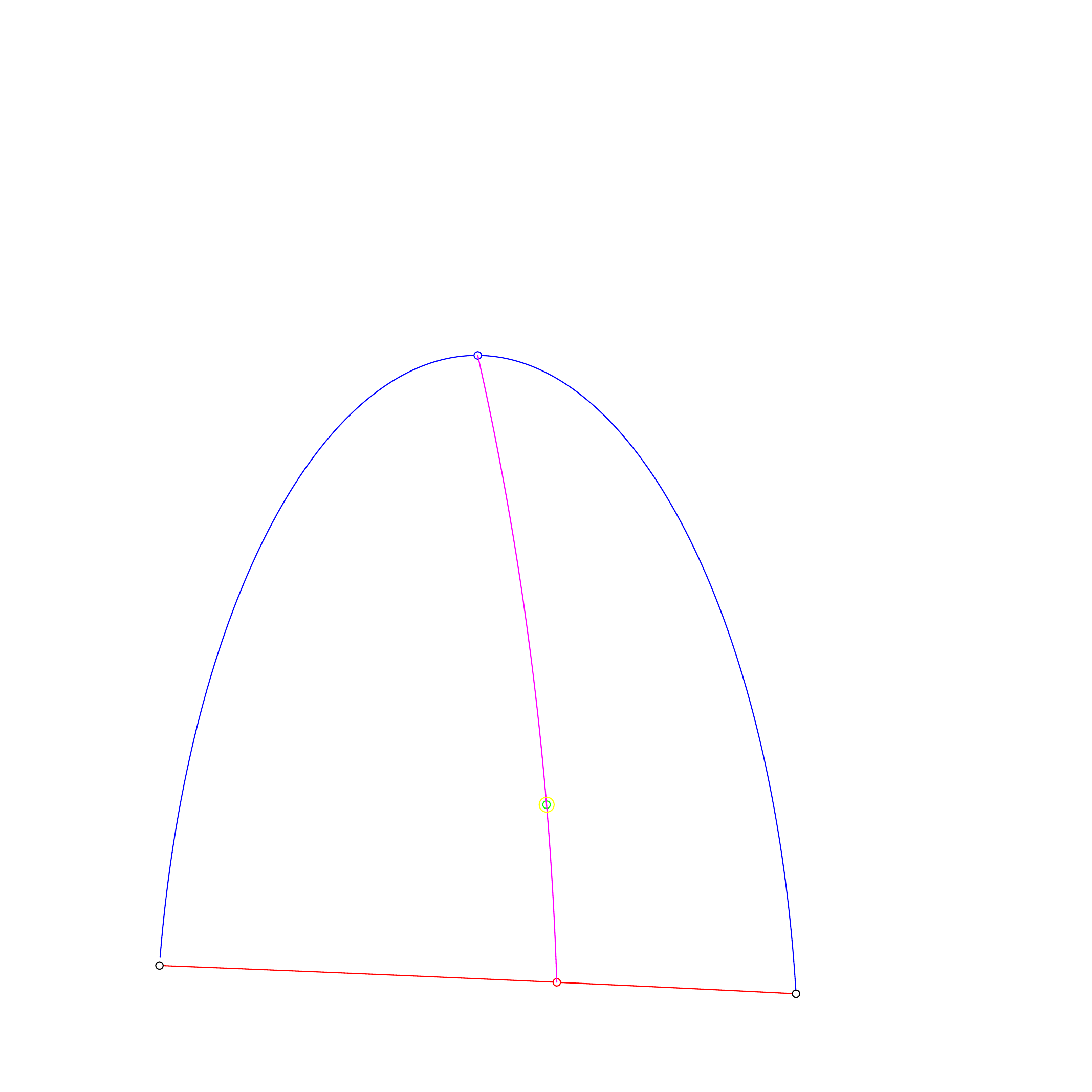}} &
\fbox{\includegraphics[width=0.35\textwidth]{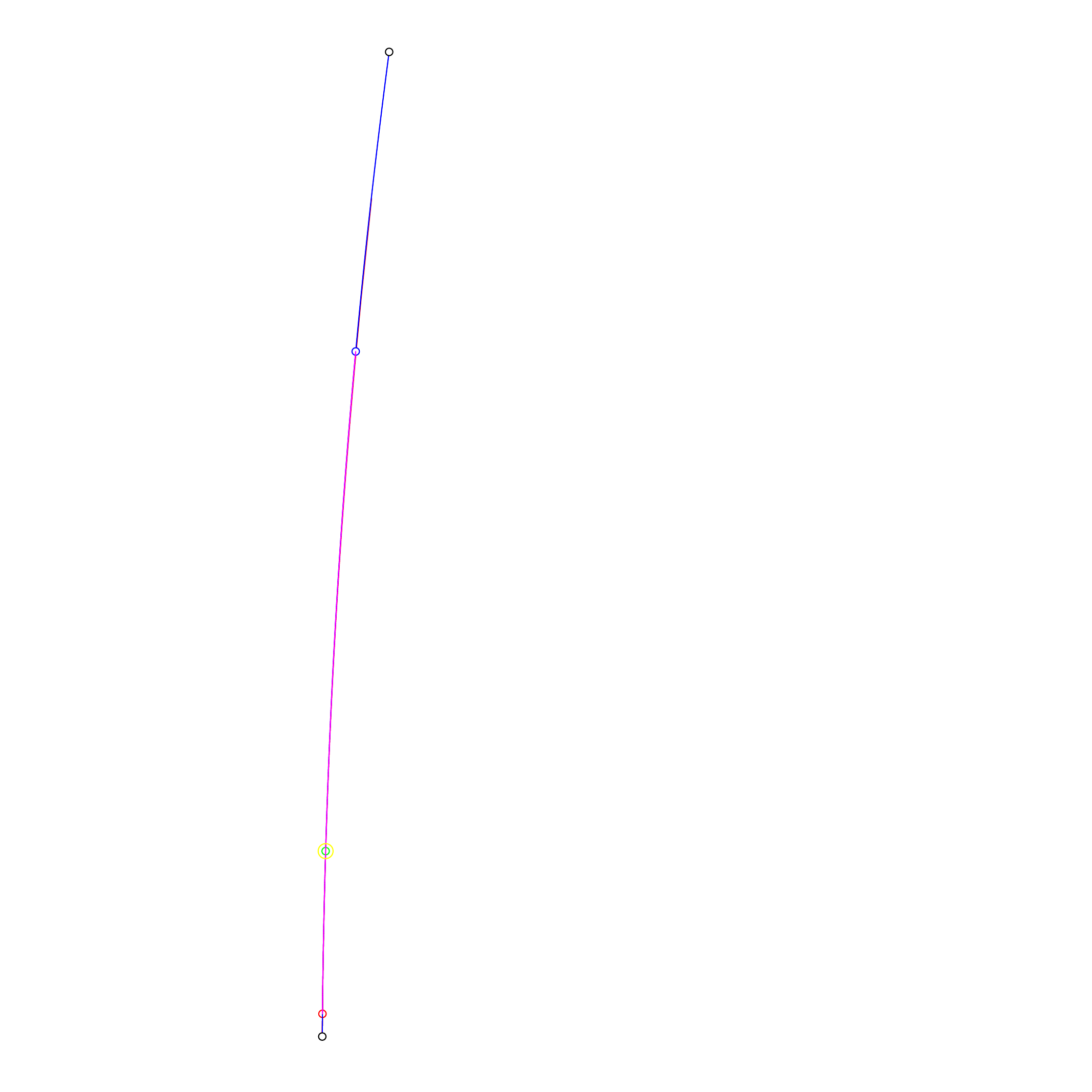}} \\
\fbox{\includegraphics[width=0.35\textwidth]{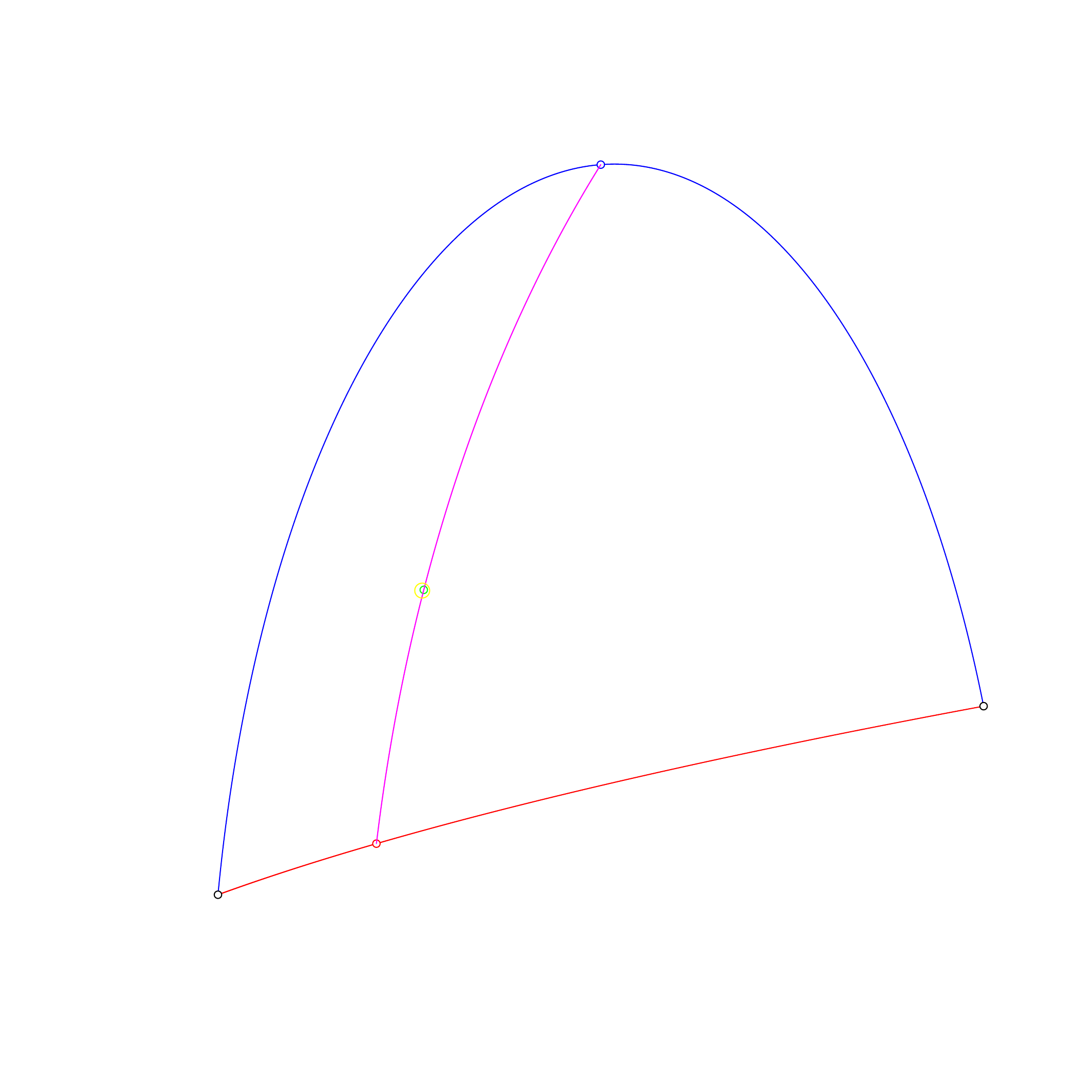}} &
\fbox{\includegraphics[width=0.35\textwidth]{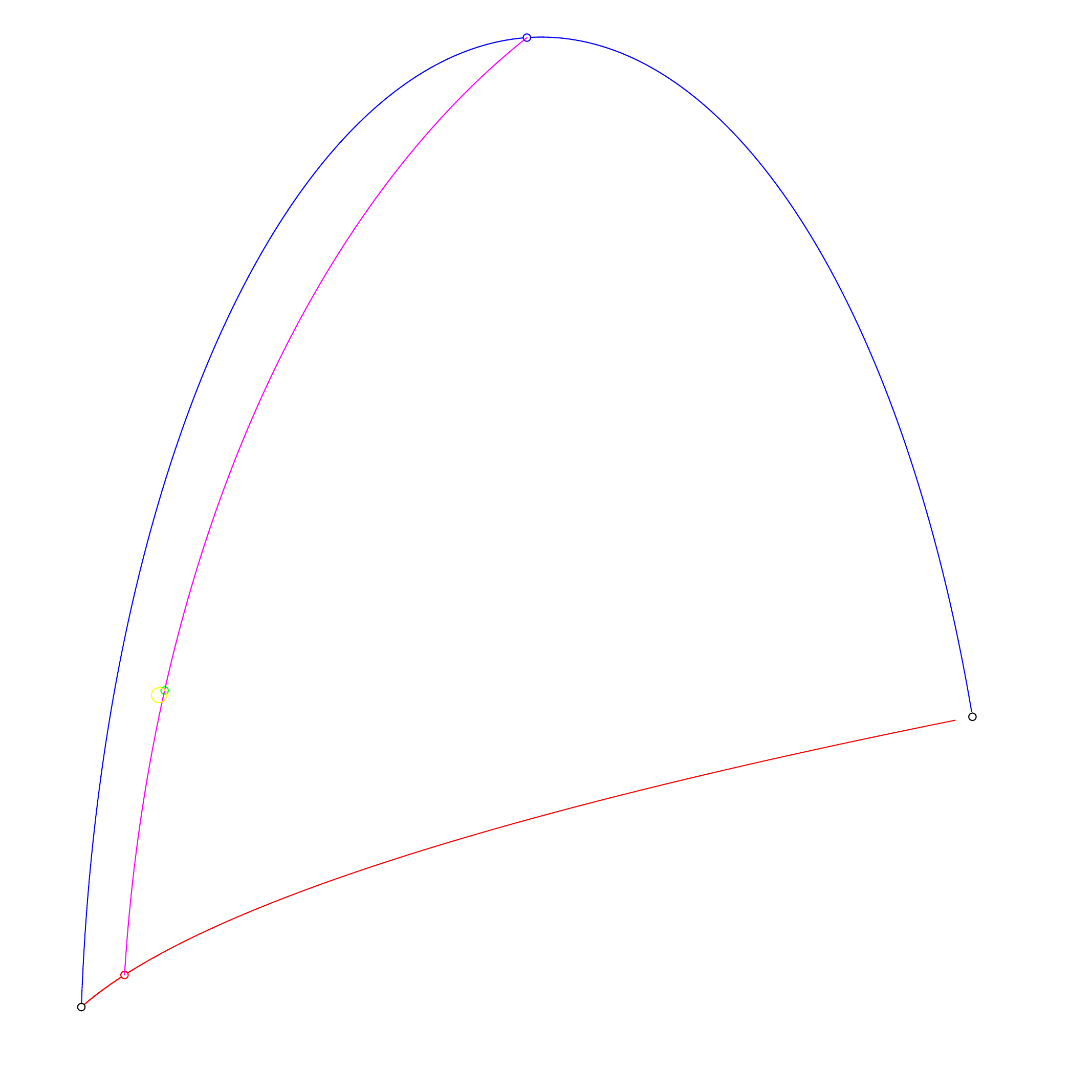}} \\
\end{tabular}
\caption{Visualization of the Jeffreys-Fisher-Rao center and Gauss-Bregman center of two univariate normal distributions (black circle).
The exponential geodesic and mixture geodesics are show in red and blue, respectively, with their corresponding midpoints.
The Jeffreys-Fisher-Rao is the Fisher-Rao midpoint lying on the Fisher-Rao geodesics (purple).
The inductive Gauss-Bregman center is displays in yellow with double size in order to ease its comparison  with the   Jeffreys-Fisher-Rao center.
}\label{fig:uninormal}
\end{figure}

Although the Rao distance between two multivariate normal distributions is not available in closed-form when $d>1$, 
the Jeffreys-Fisher-Rao center can be computed in closed-form using the closed-form geodesic with boundary condition of MVNs~\cite{kobayashi2023geodesics} 
reported in Appendix~\ref{sec:frmvnmidpoint}. Thus the Jeffreys-Fisher-Rao center is available in closed-form.
Note that the Fisher-Rao distance between normal distributions is invariant under the action of the positive affine group~\cite{calvo1990distance} as are the Jeffreys centroid, the JFR center, and the GB center.
Figure~\ref{fig:uninormal} shows several examples of the JFR and GB centers of two univariate normal distributions.
We can observe that those centers are close to each others although they are distinct when the normal distributions do not share the same means and covariance matrices. 

\begin{figure}
\centering
\begin{tabular}{cc}
\fbox{\includegraphics[width=0.45\textwidth]{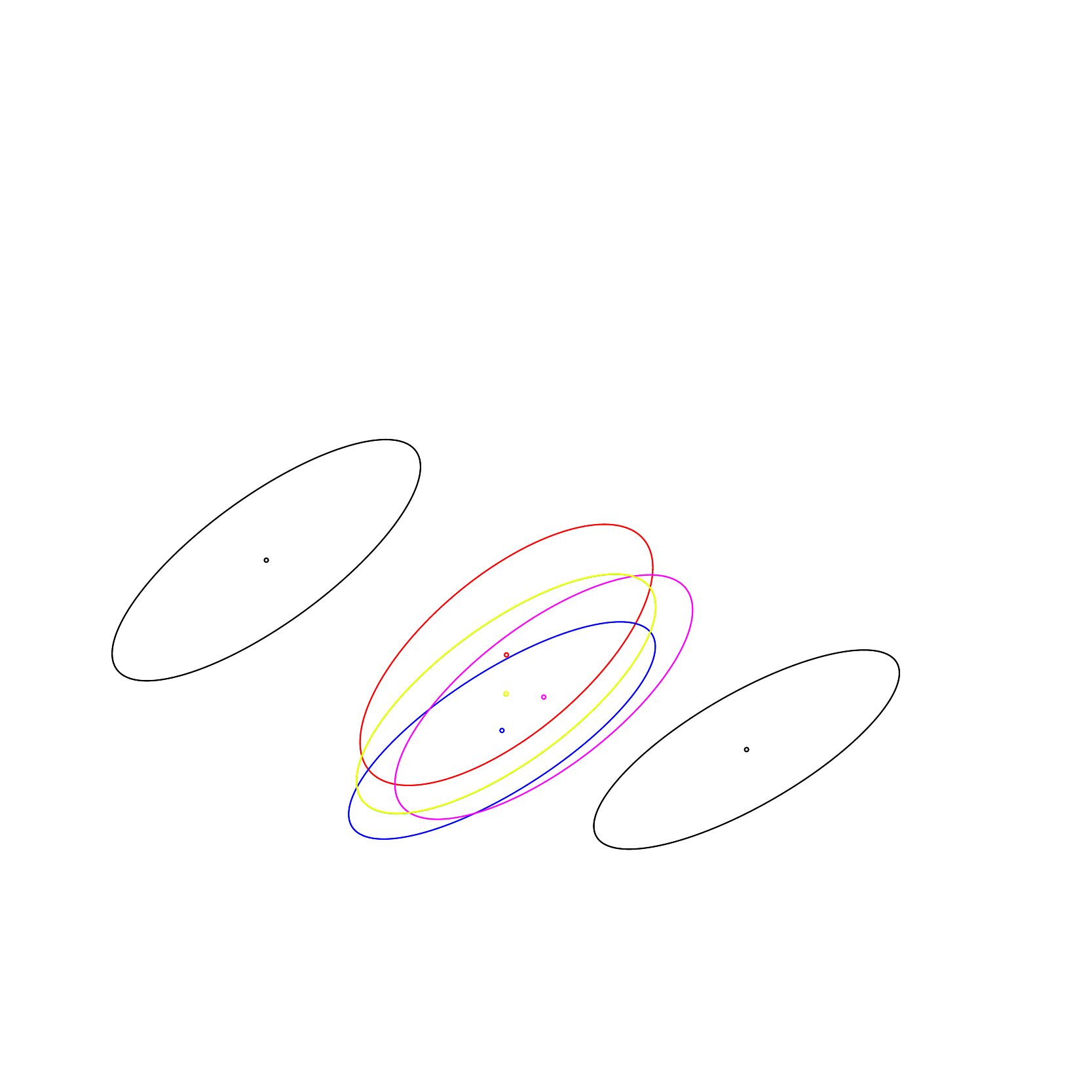}} &
\fbox{\includegraphics[width=0.45\textwidth]{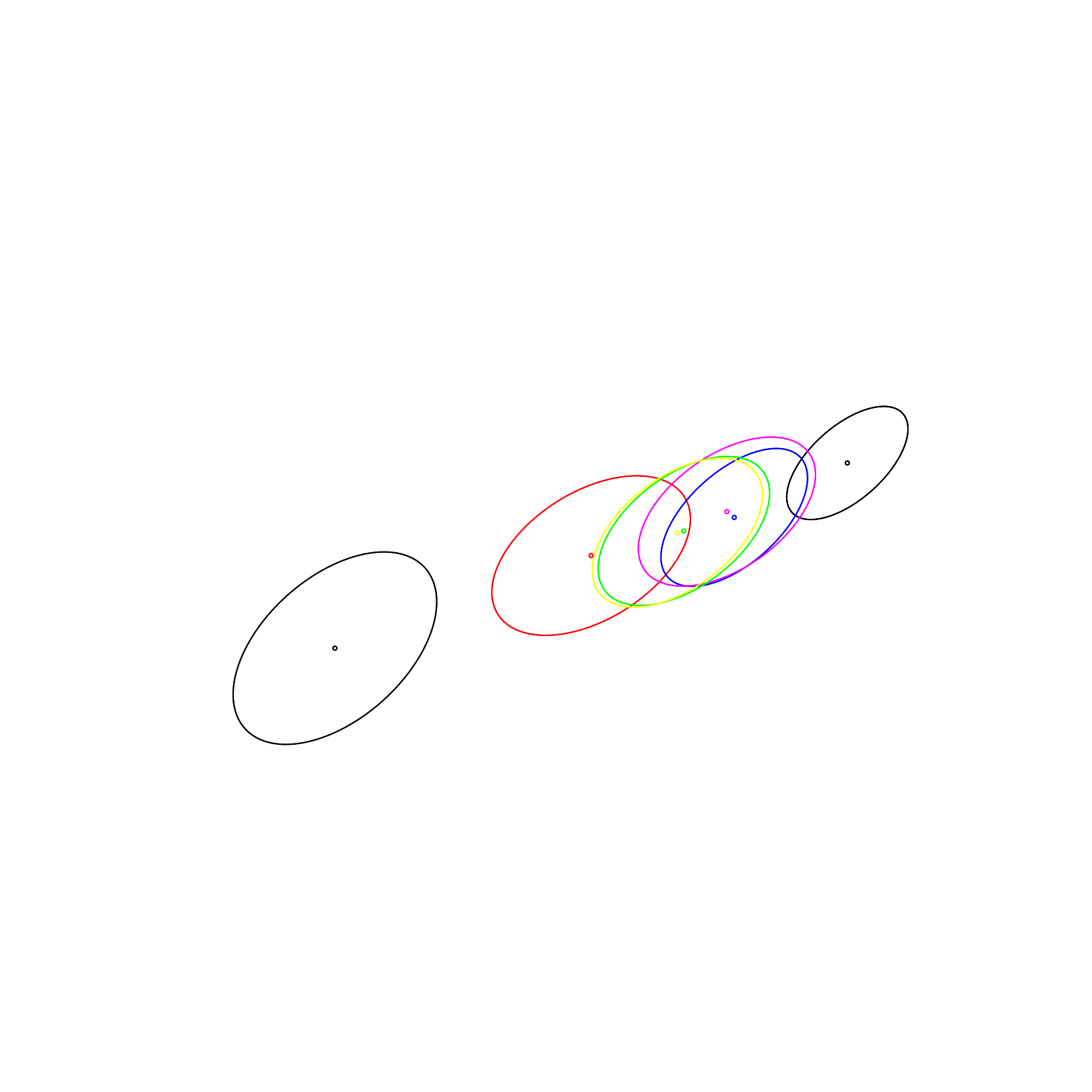}} 
\end{tabular}
\caption{Centroids and centers between a pair of bivariate normal distributions (black). Each normal distribution $N(\mu,\Sigma)$ (parameterized by a 5D parameter $\theta$) is displayed as a 2D ellipsoid 
$\calE(\mu,\Sigma)=\{(x-\mu)^\top\Sigma^{-1}(x-\mu)=l\}$ for a prescribed level $l>0$ in the sample space $\bbR^2$.}\label{fig:binormal}
\end{figure}

Figure~\ref{fig:binormal} shows the various centroids/centers between two bivariate normal distributions displayed as ellipsoids centered as their means.
Observe that the inductive Gauss-Bregman center is visually closer than the Jeffreys-Fisher-Rao center to the Jeffreys centroid.

\begin{Theorem}[JFR center of MVNs]\label{thm:jfrmvn}
The Jeffreys-Fisher-Rao center of a set of $n$ weighted multivariate normal distributions is available in closed-form. 
\end{Theorem}

\begin{figure}
\centering
\begin{tabular}{cc}
\fbox{\includegraphics[width=0.45\textwidth]{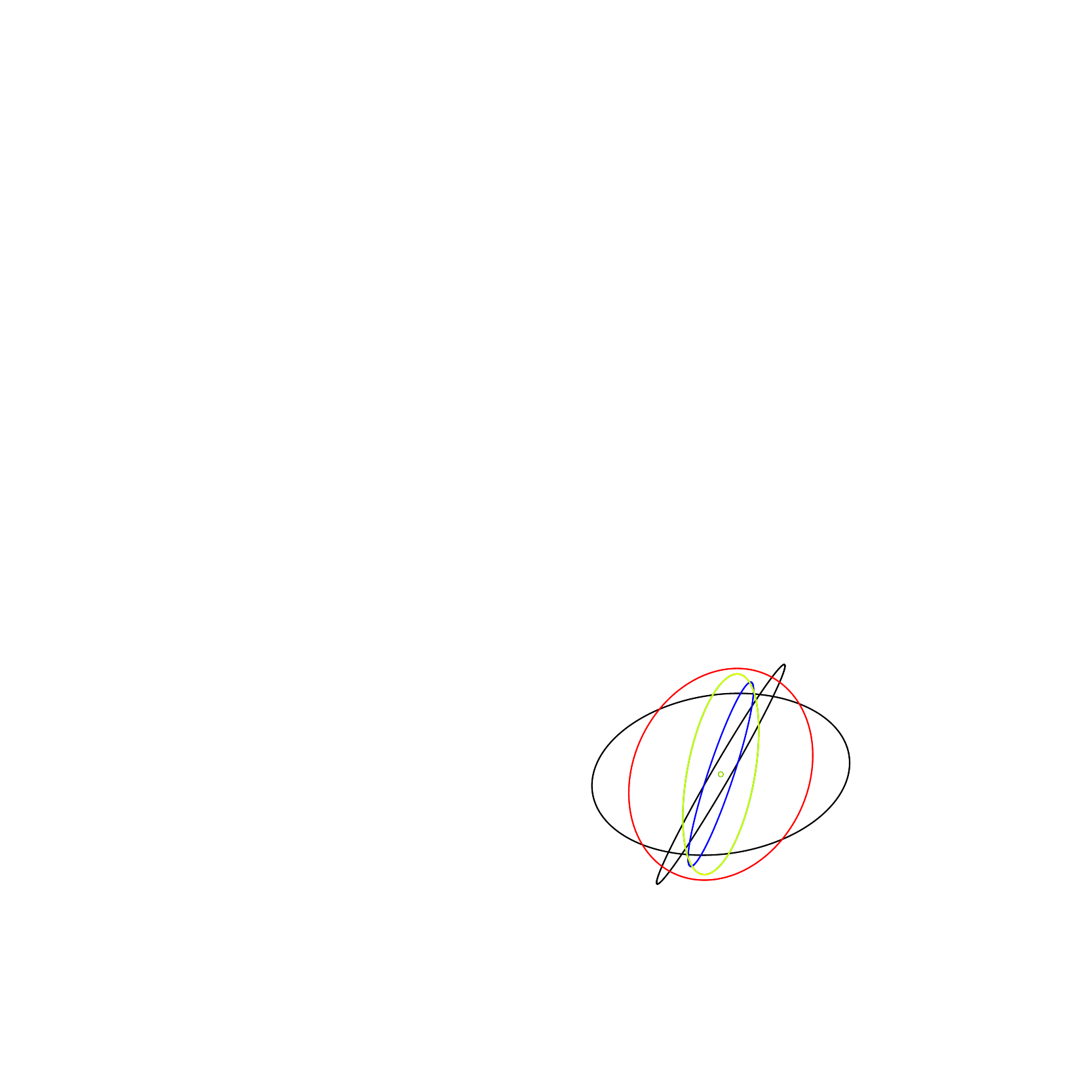}}&
\fbox{\includegraphics[width=0.45\textwidth]{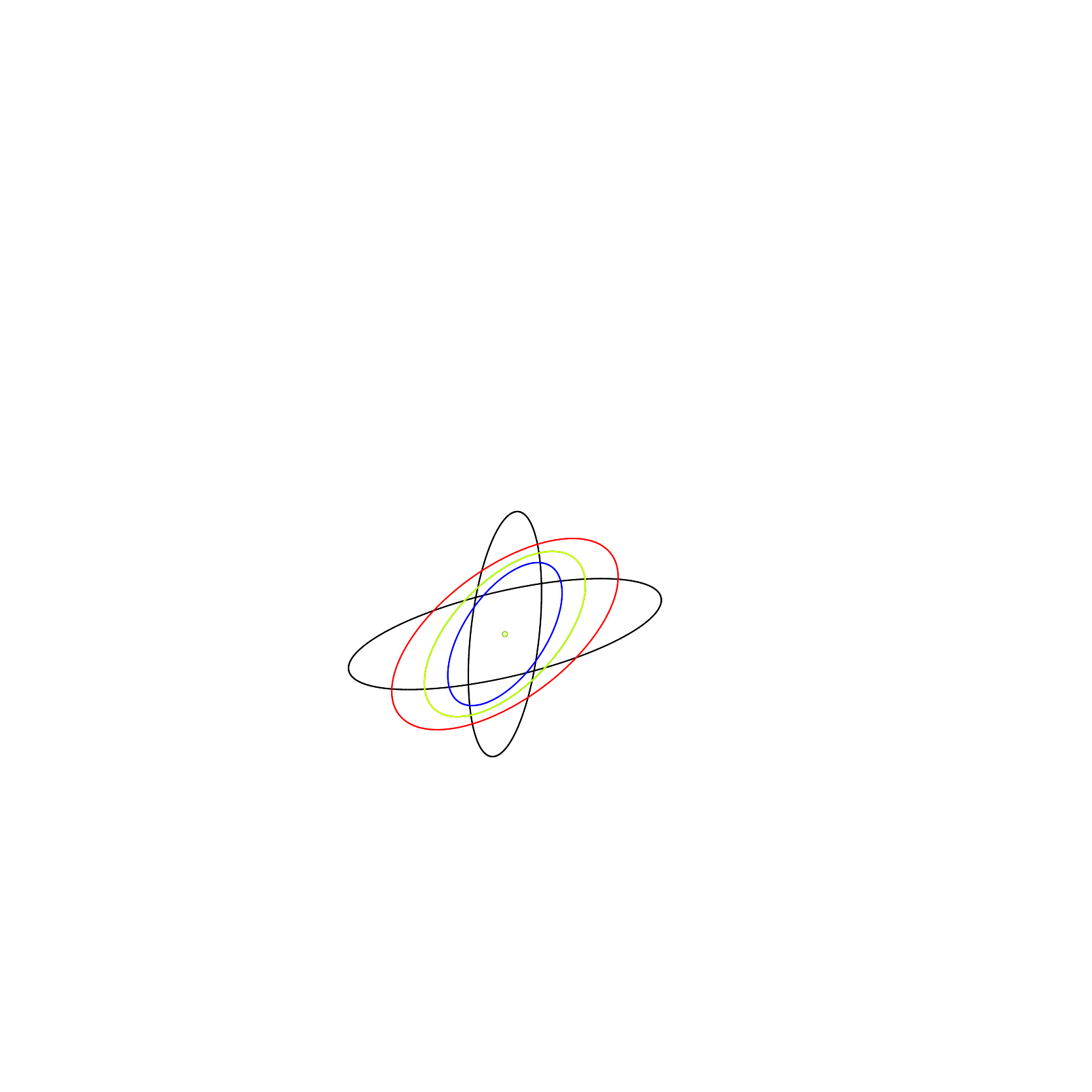}}\\
\fbox{\includegraphics[width=0.45\textwidth]{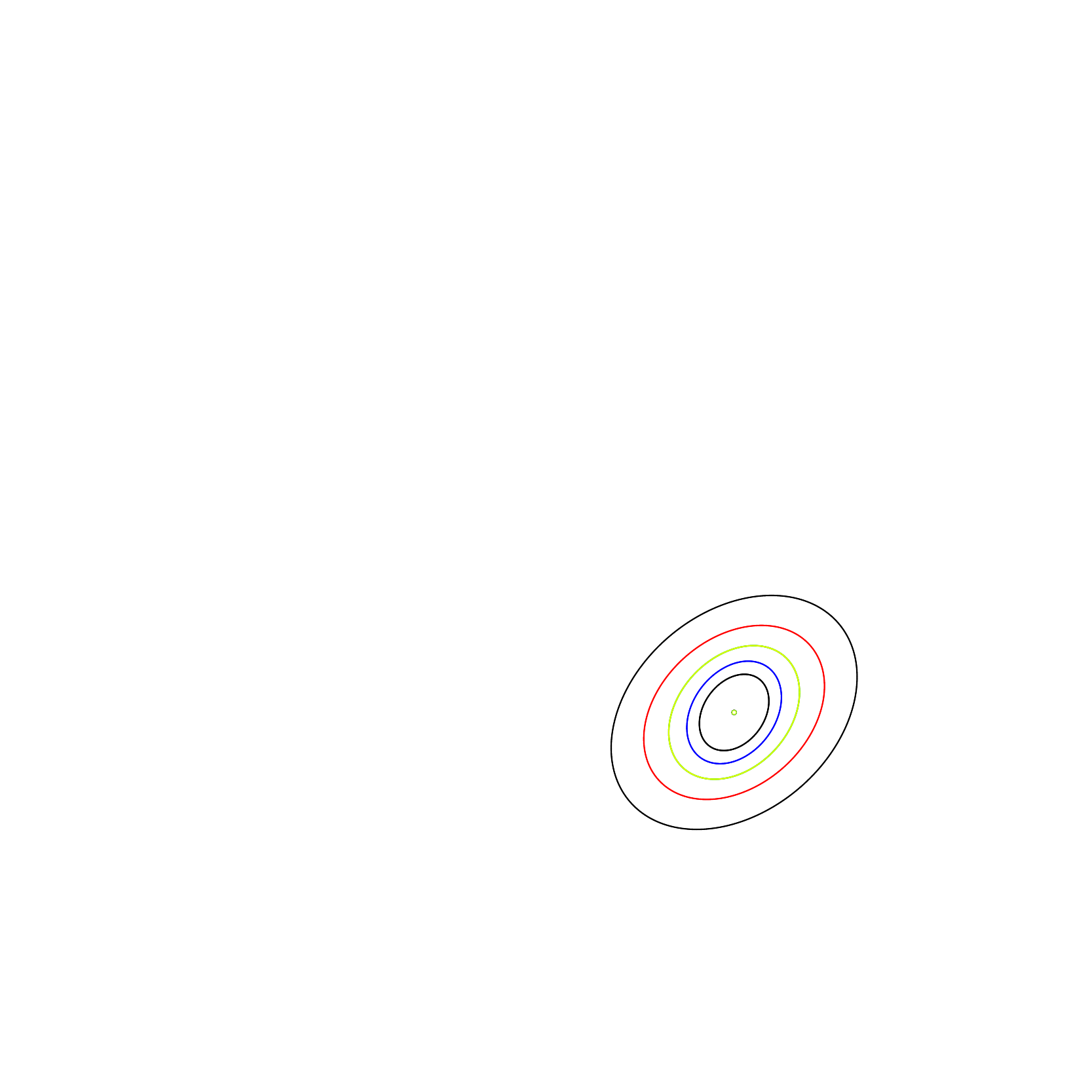}}&
\fbox{\includegraphics[width=0.45\textwidth]{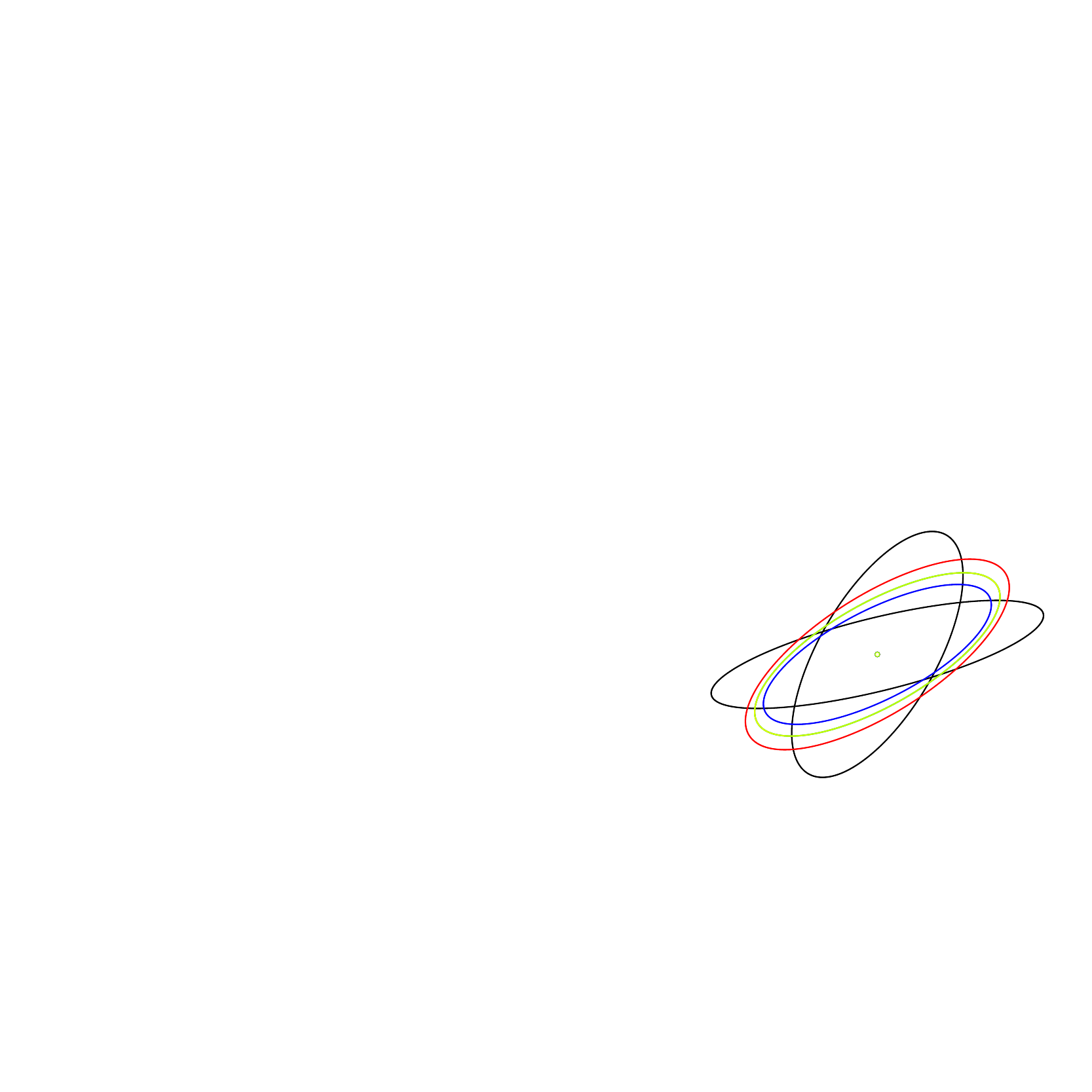}}\\  
\end{tabular}
\caption{Centroids and centers between a pair of bivariate centered normal distributions (black). Each normal distribution $N(\mu,\Sigma)$ with a prescribed $\mu$ (parameterized by a 3D parameter $\theta$) is displayed as a 2D ellipsoid. The inductive Gauss-Bregman (yellow) and Jeffreys-Fisher-Rao center (purple)  coincide perfectly in the shade of brown (purple-yellow) ellipsoids.}\label{fig:centeredbinormal}
\end{figure}

\begin{figure}
\centering
\begin{tabular}{cc}
\fbox{\includegraphics[width=0.45\textwidth]{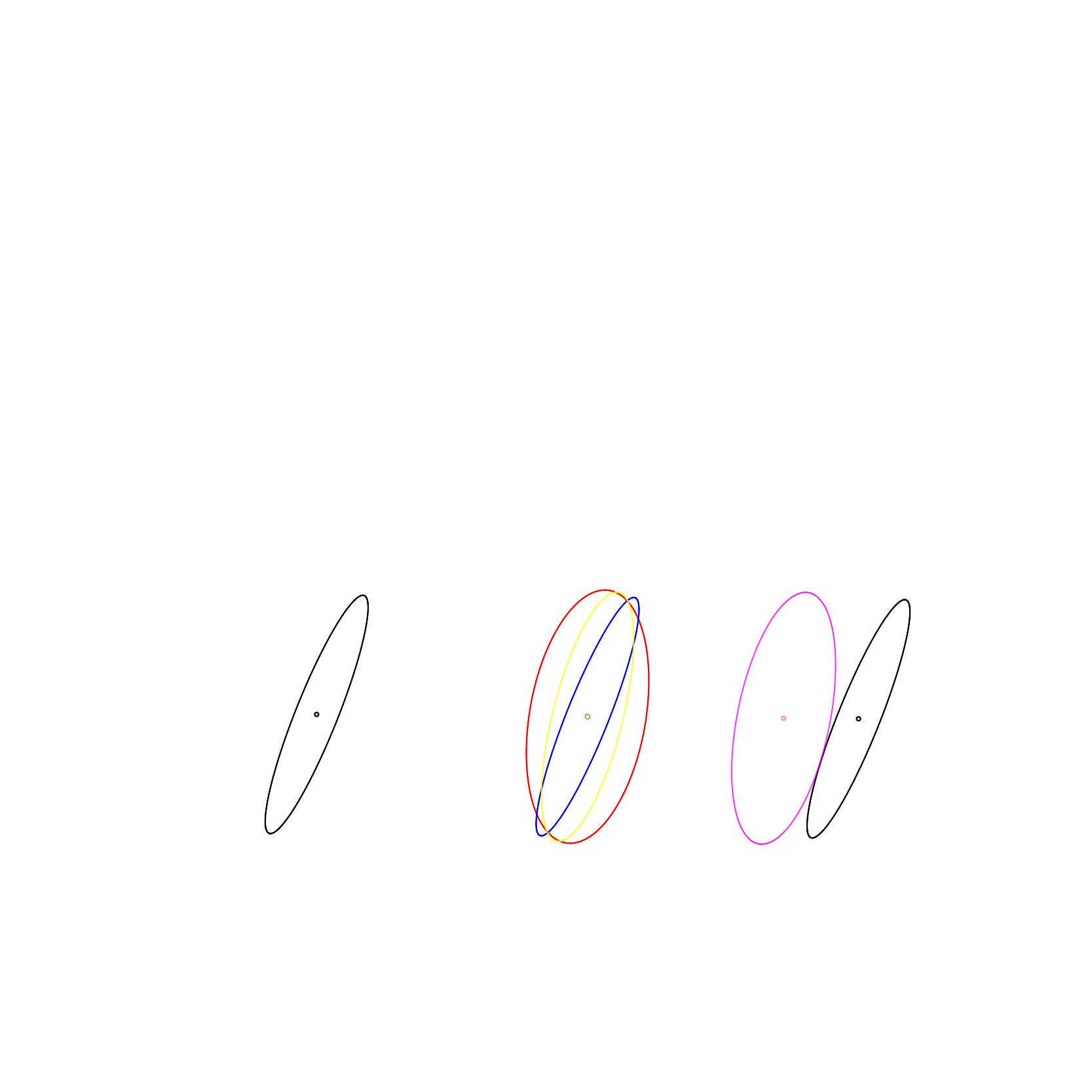}}&
\fbox{\includegraphics[width=0.45\textwidth]{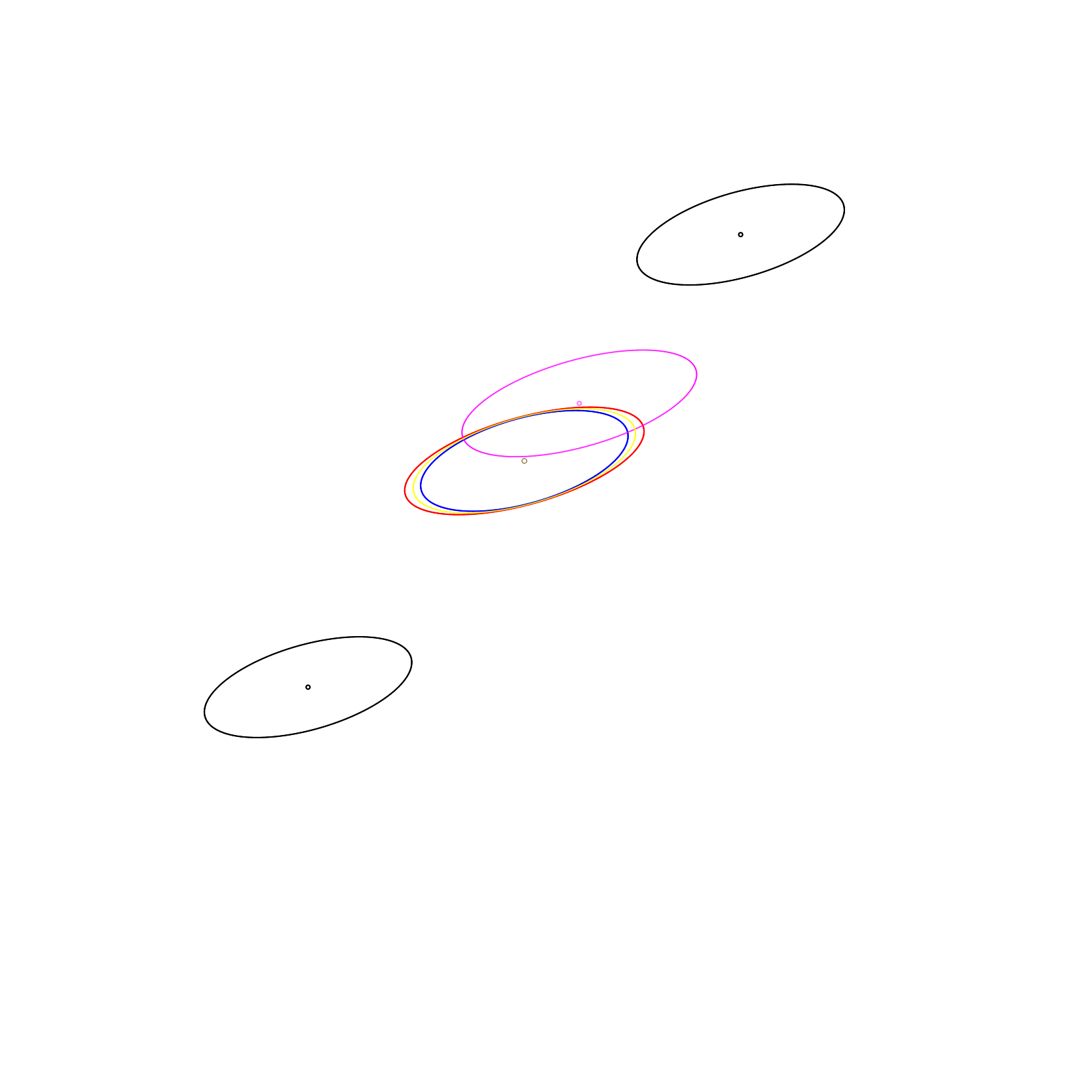}}\\
\fbox{\includegraphics[width=0.45\textwidth]{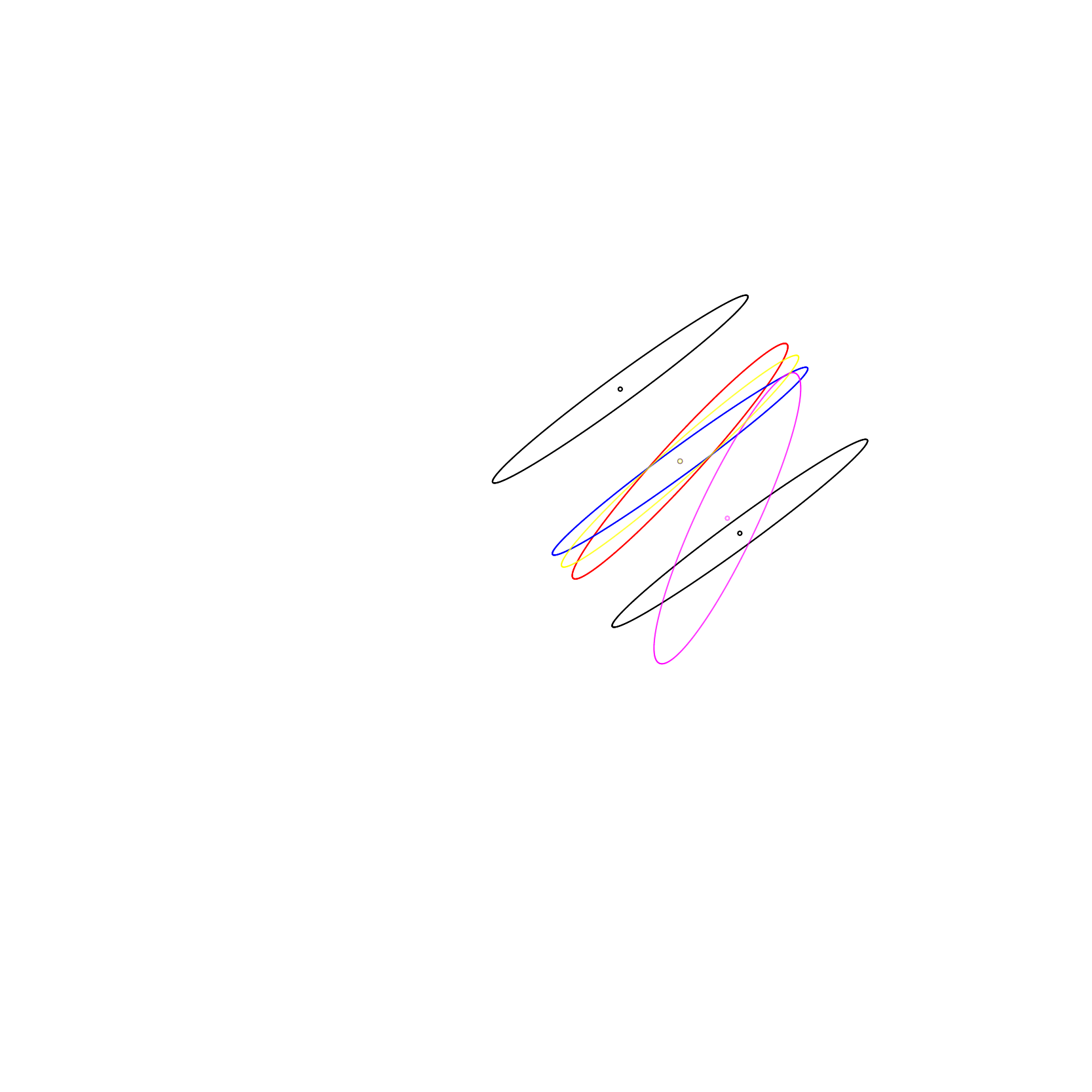}}&
\fbox{\includegraphics[width=0.45\textwidth]{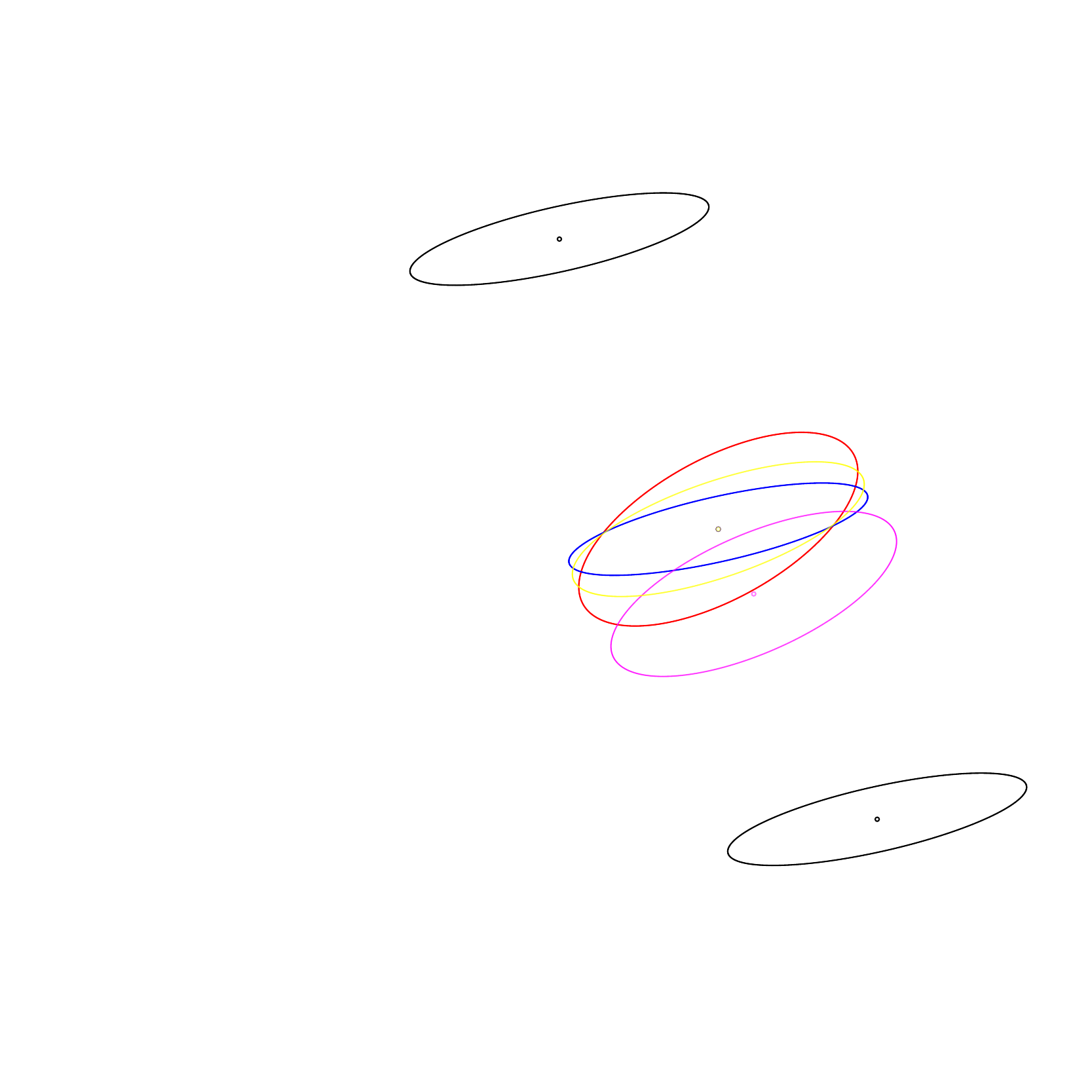}}\\  
\end{tabular}
\caption{Centroids and centers between a pair of bivariate centered normal distributions (black). Each normal distribution $N(\mu,\Sigma)$ with a prescribed covariance matrix $\Sigma$ (parameterized by a 2D parameter $\theta$) is displayed as a 2D ellipsoid. The inductive Gauss-Bregman (yellow) and Jeffreys-Fisher-Rao center (purple)  do not coincide.}\label{fig:samecovarbinormal}
\end{figure}

Figure~\ref{fig:centeredbinormal} displays the various centroids and centers for  pairs of bivariate normal distributions centered at the same mean.
Figure~\ref{fig:samecovarbinormal} shows the centroids and centers for  pairs of bivariate normal distributions with the same covariance matrix.

\begin{Remark}
In general, an exponential family may be characterized equivalently by two convex functions: 
(1) its log-normalizer $F(\theta)$ or (2) its partition function $Z(\theta)=\exp(F(\theta))$ which is log-convex and hence also convex~\cite{nielsen2024divergences}.
It has been shown that the Bregman divergence $B_Z$ for $Z=\sqrt{\det(\theta)}$ (convex) corresponds to the reverse extended Kullback-Leibler divergence between unnormalized PDFs of normal distributions:
$$
B_Z(\theta_1:\theta_2)=D_\KL^+(\tilde{p}_{\lambda(\theta_2)}:\tilde{p}_{\lambda(\theta_1)}),
$$
where $\tilde{p}_{\mu,\Sigma}=\exp\left(-\frac{1}{2}(x-\mu)^\top\Sigma^{-1}(x-\mu)\right)$ and the extended KLD between two positive measures is given by:
$$
D_\KL^+(m_1:m_2)=\int \left( m_1(x)\log\frac{m_1(x)}{m_2(x)}+m_2(x)-m_1(x) \right) \dmu(x).
$$
\end{Remark}

\begin{Remark}
We may further define yet another center for multivariate normal distributions by considering the Fisher-Rao isometric embedding of the Fisher-Rao $d$-variate normal manifold $\calM=\{p_{\mu,\Sigma}\}$ into the Fisher-Rao $(d+1)$-variate centered manifold $\calN_0^+=\{q_P(y)=p_{0,P}(y) \st P\in\Sym^{++}(\bbR,d+1)\}$ using Calvo \& Oller mapping~\cite{calvo1990distance}:
$$
f(\mu,\Sigma) \eqdef
  \mattwotwo{\Sigma+\mu\mu^\top}{\mu}{\mu^\top}{1}.
$$

Let $\bar\calM=\{f(p)\st p\in\calM\}$ denote the embedded submanifold of codimension $1$ in $\calN_0^+$.
The Calvo-Oller center is then defined by taking the Fisher-Rao midpoints $q_\CO$ of $q_{P_1}$ and $q_{P_2}$,  projecting $q_{\CO}$ onto $\bar\calM$ as $q_\CO'$, and converting $q_\CO'$ into $p_\CO\in\calM$ using the inverse mapping $f^{-1}$~\cite{nielsen2023simple}.

The Fisher orthogonal projection of a $(d+1)\times(d+1)$ matrix $P\in\calN_0^+$ onto the submanifold $\bar\calM$  is done as follows:
Let $\beta=P_{d+1,d+1}$ and write  $P=\mattwotwo{\Sigma+\beta\mu\mu^\top}{\beta\mu}{\beta\mu^\top}{\beta}$.
Then the orthogonal projection at $P\in\calP$ onto $\bar\calM$ is $\mattwotwo{\Sigma+\mu\mu^\top}{\mu^\top}{\mu}{1}$.
 See~\cite{nielsen2023simple} for details of the Calvo \& Oller embedding/projection method.
\end{Remark}

\section{Experiments}\label{sec:experiments}

We run all experiments on a Dell Inspiron 5502 Core i7-116567@2.8Ghz using compiled Java programs.
For each experiment, we consider a set of $n=2$ uniformly randomized histograms with $d$ bins (i.e., points in $\Delta_d$) and calculate the numerical Jeffreys centroid which requires the time-consuming Lambert $W$ function, the GB center and the JFR center.
For each prescribed value of $d$, we run $10000$ experiments to collect various statistics like the average and maximum approximations and running times.
The approximations of the JFR and GB methods are calculated either as the approximation of the Jeffreys information (Eq.~\ref{eq:infoJ}) or as the approximation of the centers with respect to the numerical Jeffreys centroids measured using the total variation distance.
Table~\ref{tab:exp} is a verbatim export of our experimental results when we range the dimension of histograms for $d=2$ to $d=256$, by doubling the dimension at each round.
The inductive GB center is stopped when the total variation $\frac{1}{2}\|a_t-g_t\|_1\leq 10^{-8}$.

\begin{sidewaystable}
{\scalebox{0.85}{
\begin{tabular}{l|llllll|llllll}
dim. &  \multicolumn{6}{c}{Jeffreys-Fisher-Rao center} & \multicolumn{6}{c}{Gauss-Bregman center}\\
&   avg info $\eps$ &  max info $\eps$ &  avg TV & max TV & avg time &    $\times$ speed  &   avg info $\eps$ &  max info $\eps$ &    avg TV & max TV & avg time &    $\times$ speed \\ \hline
d=2 & 5.662e-06 & 6.386e-03 & 8.735e-05  & 5.005e-02 & 1.614e-07 & 82.541 & 1.507e-04 & 9.745e-02 & 6.304e-04 &  5.005e-02 & 5.072e-07 & 26.258 \\
d=4 & 1.283e-05 & 5.294e-03 & 1.690e-04  & 3.969e-02 & 1.418e-07 & 182.309 & 4.696e-04 & 7.695e-02 & 1.431e-03 &  3.969e-02 & 1.623e-07 & 159.304 \\
d=8 & 2.766e-05 & 6.970e-03 & 2.210e-04  & 3.470e-02 & 1.772e-07 & 292.125 & 1.011e-03 & 9.677e-02 & 2.033e-03 &  3.470e-02 & 1.955e-07 & 264.680 \\
d=16 & 3.531e-05 & 8.544e-03 & 2.325e-04  & 2.450e-02 & 6.318e-07 & 224.370 & 1.388e-03 & 9.231e-02 & 2.275e-03 &  2.450e-02 & 7.208e-07 & 196.660 \\
d=32 & 4.123e-05 & 5.242e-03 & 2.457e-04  & 1.230e-02 & 4.811e-07 & 462.754 & 1.674e-03 & 5.398e-02 & 2.449e-03 &  1.230e-02 & 5.457e-07 & 408.007 \\
d=64 & 4.747e-05 & 3.437e-03 & 2.486e-04  & 9.756e-03 & 9.789e-07 & 578.354 & 1.863e-03 & 3.685e-02 & 2.498e-03 &  9.756e-03 & 1.160e-06 & 488.246 \\
d=128 & 5.020e-05 & 2.540e-03 & 2.491e-04  & 6.580e-03 & 5.874e-06 & 477.412 & 1.937e-03 & 2.374e-02 & 2.522e-03 &  6.580e-03 & 6.605e-06 & 424.609 \\
d=256 & 4.735e-05 & 1.410e-03 & 2.476e-04  & 4.855e-03 & 9.349e-06 & 528.452 & 1.914e-03 & 1.521e-02 & 2.529e-03 &  4.855e-03 & 1.110e-05 & 445.304 
\end{tabular}
} 
}
\caption{Experiments for JFR and GB centers approximating the numerical Jeffreys centroid.}\label{tab:exp}
\end{sidewaystable}

We observe that the JFR center is faster to compute than the GB center but the GB center is of higher quality (i.e., better approximation with lower $\eps$) than the JFR center to approximate the numerical Jeffreys centroid.

Another test consists in choosing $d=3$ and the following  two 3D normalized histograms:
 $(\frac{1}{3},\frac{1}{3},\frac{1}{3})$ and $(1-\alpha,\alpha/2,\alpha/2)$
 for $\alpha\in\{10^{-1},10^{-2},\ldots,10^{-7},10^{-8}\}$.
Table~\ref{tab:exp2} reports the experiments. The objective is to find a setting where the both the JFR and GB centers are distinguished from the Jeffreys centroid. 
We see that as we decrease $\alpha$, the approximation factor $\epsilon$ gets worse for both the JFR center and the GB center.
The JFR center is often faster to compute than the GB inductive center, but the approximation of the GB center is better than the JFR approximation.

\begin{table}
\centering
{\scalebox{0.85}{
\begin{tabular}{l|llll|llll}
$\alpha$ &   Info. $\eps$ &   TV $\eps$ &    avg time &    $\times$ speed  &  Info.   $\eps$ &  TV $\eps$ &  avg time &    $\times$ speed \\ \hline
1.000e-01 & 6.882e-09 &   2.495e-05 & 1.767e-07  & 125.960 & 1.338e-06 &  3.480e-04 & 2.334e-07 &  95.356  \\
1.000e-02 & 2.607e-05 &   1.722e-03 & 1.371e-07  & 167.932 & 1.061e-03 &  1.108e-02 & 1.565e-07 &  147.104  \\
1.000e-03 & 6.262e-04 &   7.530e-03 & 1.033e-07  & 218.450 & 1.272e-02 &  3.534e-02 & 1.208e-07 &  186.698  \\
1.000e-04 & 3.632e-03 &   1.570e-02 & 1.171e-07  & 193.345 & 4.580e-02 &  6.065e-02 & 1.367e-07 &  165.571  \\
1.000e-05 & 1.121e-02 &   2.419e-02 & 1.546e-07  & 150.807 & 7.322e-03 &  1.929e-02 & 2.834e-07 &  82.261  \\
1.000e-06 & 2.457e-02 &   3.204e-02 & 1.619e-07  & 141.896 & 1.655e-02 &  2.579e-02 & 2.512e-07 &  91.467  \\
1.000e-07 & 4.375e-02 &   3.897e-02 & 1.357e-07  & 170.065 & 3.065e-02 &  3.183e-02 & 2.131e-07 &  108.314  \\
1.000e-08 & 6.806e-02 &   4.492e-02 & 1.315e-07  & 173.698 & 4.948e-02 &  3.725e-02 & 2.017e-07 &  113.292  \\
1.000e-09 & 9.651e-02 &   4.999e-02 & 1.125e-07  & 208.627 & 7.240e-02 &  4.199e-02 & 1.590e-07 &  147.610  \\
1.000e-10 & 1.281e-01 &   5.428e-02 & 8.366e-08  & 242.967 & 9.862e-02 &  4.610e-02 & 1.111e-07 &  183.000  \\
1.000e-11 & 1.621e-01 &   5.792e-02 & 1.066e-07  & 215.817 & 1.274e-01 &  4.963e-02 & 1.325e-07 &  173.614  \\
1.000e-12 & 1.979e-01 &   6.100e-02 & 1.028e-07  & 229.484 & 1.580e-01 &  5.266e-02 & 1.329e-07 &  177.581  \\
1.000e-13 & 2.348e-01 &   6.363e-02 & 9.541e-08  & 244.587 & 1.901e-01 &  5.526e-02 & 1.255e-07 &  185.940  \\
1.000e-14 & 2.727e-01 &   6.589e-02 & 1.062e-07  & 219.787 & 2.231e-01 &  5.750e-02 & 1.361e-07 &  171.456  \\
1.000e-15 & 3.112e-01 &   6.784e-02 & 9.043e-08  & 248.688 & 2.570e-01 &  5.943e-02 & 1.322e-07 &  170.122  \\
1.000e-16 & 3.483e-01 &   6.946e-02 & 8.857e-08  & 267.219 & 2.897e-01 &  6.105e-02 & 1.438e-07 &  164.535  
\end{tabular}
}
}
\caption{Experiments for JFR and GB centers approximating the numerical Jeffreys centroid for the following setting of two normalized histograms of $3$ bins: $(\frac{1}{3},\frac{1}{3},\frac{1}{3})$ and $(1-\alpha,\alpha/2,\alpha/2)$.}\label{tab:exp2}
\end{table}

\section{Conclusion and discussion}\label{sec:concl}

\begin{table}
\centering
\begin{tabular}{lccc}
Family & Jeffreys & Jeffreys-Fisher-Rao & Gauss-Bregman \\ \hline\hline
Exponential family       &    Eq.~\ref{eq:Jeffreyscentroid}      &   Definition~\ref{def:JFR}                   & Definition~\ref{def:GBcenter}\\  \hline
1d Exponential family & $\times$ & $\triangle$ & $\times$ \\
                      &     Theorem~\ref{thm:Jef1d}      &   & \cite{lehmer1971compounding} \\ \hline
Categorical family & $\triangle$ & $\surd$ & $\times$\\
                   &  \cite{nielsen2013jeffreys} & Theorem~\ref{thm:Jcat} & Theorem~\ref{thm:convergence}\\ \hline
Normal family & $\times$ & $\surd$ & $\times$\\
              &   \cite{nielsen2009sided} & Theorem~\ref{thm:jfrmvn} &  Theorem~\ref{thm:convergence} \\ \hline
Centered normal family & $\surd$ & $\surd$ & $\surd$\\
                       & \cite{moakher2006symmetric} & \cite{moakher2006symmetric} & \cite{nakamura2001algorithms}\\ \hline
\end{tabular}
\caption{Summary of the results: $\triangle$ indicates a generic formula, $\surd$ a closed-form formula and $\times$ no-known formula.}\label{tab:results}
\end{table}

In this work, we considered the Jeffreys centroid of a finite weighted set of densities of a given exponential family $\calE=\{p_\theta(x)\st\theta\in\Theta\}$.
This Jeffreys centroid amounts to a symmetrized Bregman centroid on the corresponding weighted set of natural parameters of the densities~\cite{nielsen2009sided}. 
In general,  the Jeffreys centroids does not admit closed-form formula~\cite{nielsen2009sided,nielsen2013jeffreys} except for sets of same-mean   normal distributions~\cite{moakher2006symmetric} (see Appendix~\ref{sec:logdet}).

In this paper, we interpreted the closed-form formula for same-mean normal distributions in two different ways: 
\begin{itemize}
\item First, as the Fisher-Rao geodesic midpoint of the sided Kullback-Leibler centroids. This interpretation lets us relax the midpoint definition to arbitrary exponential families to define the Jeffreys-Fisher-Rao center (JFR center of Definition~\ref{def:JFR}), and

\item Second, as an inductive
 $(A,m_{\nabla F})$-center using a multivariate Gauss-type double sequence which converges to the Gauss-Bregman center (GB center of Definition~\ref{def:GBcenter}).
The latter definition yields an extension of Nakamura's arithmetic-harmonic $(A,H)$-mean~\cite{nakamura2001algorithms} to arbitrary $(A,m_{\nabla F})$-mean for which we proved convergence under a separability condition in Theorem~\ref{thm:convergence}. Convergence proof remains to be done in the general case although we noticed in practice convergence when $\nabla F(\theta)$ is the moment parameter of categorical or normal distributions.
\end{itemize}

In general, the Jeffreys, JFR, and GB centers differ from each others (e.g., the case of categorical distributions).
But for sets of same-mean  normal distributions, they remarkably coincide altogether: Namely, this was the point of departure of this research.
We reported generic or closed-form formula for the JFR centers of (a) uni-order parametric families in \S\ref{sec:JFRuni} (Theorem~\ref{thm:Jef1d}), 
(b) categorical families in \S\ref{sec:JFRcat} (Theorem~\ref{thm:Jcat}), and (c) multivariate normal families in \S\ref{sec:JFRmvn} (Theorem~\ref{thm:jfrmvn}).
Table~\ref{tab:results} summarizes the new results obtained in this paper and state references of prior work.
Notice that in practice, we approximate the Gauss-Bregman center by prescribing a number of iterations $T\in\bbN$ for the Gauss-Bregman double sequence to get $m_\GB^{(T)}$.
Prescribing the number of GB center iterations $T$ allows us to tune the time complexity of computing $m_\GB^{(T)}$ while adjusting the quality of the approximation of the Jeffreys centroid.

In applications requiring Jeffreys centroid, we thus propose to either use the fast  Jeffreys-Fisher-Rao center when a closed-form formula is available for the family of distributions at hand, or use the Gauss-Bregman center approximation with a prescribed number of iterations as a drop-in replacement of the numerical Jeffreys centroids while keeping the Jeffreys divergence. (The centers we defined are not centroids as we do not exhibit distances from which they are population minimizers.)
 
\begin{figure}
\centering
\includegraphics[width=0.65\textwidth]{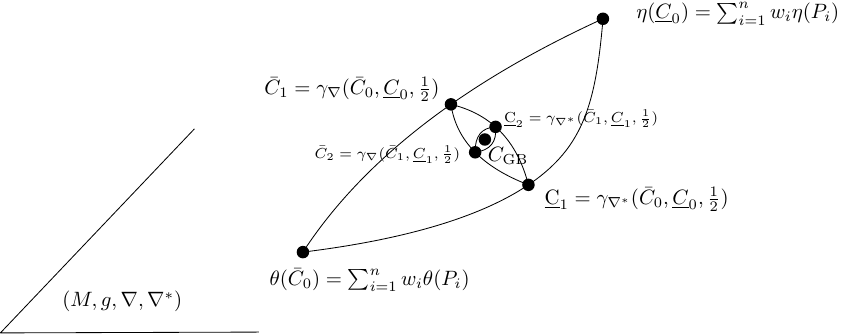}
\caption{Illustration on a dually flat space of the double sequence inducing a Gauss-Bregman center in the limit.}\label{fig:GBDFS}
\end{figure}

More generally, let us  rephrase the results in a purely geometric setting using the framework of information geometry~\cite{IG-2016}:
Let $P_1,\ldots, P_n$ be a set of $n$ points weighted by a vector $w\in\Delta_n$ on a $m$-dimensional dually flat space $(M,g,\nabla,\nabla^*)$ with $\nabla$-affine coordinate system $\theta(\cdot)$ and dual $\nabla^*$-affine coordinate system $\eta(\cdot)$, where $\nabla$ and $\nabla^*$ are two torsion-free dual affine connections. The Riemannian metric $g$ is a Hessian metric~\cite{shima2007geometry} which may be expressed in the $\theta$-coordinate system as $g(\theta)=\nabla^2 F(\theta)$ or in the dual coordinate system as $g(\eta)=\nabla^2 F^*(\eta)$ where $F(\theta)$ and $F^*(\eta)$ are dual convex potential functions related by the Legendre-Fenchel transform~\cite{IG-2016,shima2007geometry}. 
Let $\eta_i=\nabla F(\theta_i)$ and $\theta_i=\nabla F^*(\eta_i)$ and  be the coordinates of point $P_i$ in the $\eta$- and $\theta$-coordinate systems, respectively.
An arbitrary point $P$ can be either referenced in the $\theta$-coordinate system ($P=P_\theta$) or in the $\eta$-coordinate system ($P=P_\eta$).
Then the Jeffreys-Fisher-Rao center is defined as the midpoint with respect to the Levi-Civita connection $\bar\nabla=\frac{\nabla+\nabla^*}{2}=\nabla^g$ of $g$
\begin{equation}
C_{\JFR} := \gamma_{\bar\nabla}(C_{\bartheta},C_{\ubartheta},\frac{1}{2}) =: C_{\bartheta} \# C_{\ubartheta}.
\end{equation}
The point $C_{\bartheta}$ is the centroid with respect to the canonical flat divergence $\calD(P:Q)=F(\theta(P))+F^*(\eta(Q))-\sum_{i=1}^m\theta_i(P)\eta_i(Q)$, and the point $C_{\ubartheta}$ is the centroid with respect to the dual canonical flat divergence $\calD^*(P:Q):=\calD(Q:P)$.
The canonical divergence is expressed using the mixed coordinates $\theta$/$\eta$, but can also be expressed using the $\theta$-coordinates as an equivalent Bregman divergence $\calD(P:Q)=B_F(\theta(P):\theta(Q))$ or as a reverse dual Bregman divergence $\calD(P:Q)=B_{F^*}(\eta(Q):\eta(P))$.
This JFR center $C_{\JFR}$ approximates the symmetrized centroid with respect to the canonical symmetrized divergence $\calS(P,Q)=\calD(P:Q)+\calD(Q:P)$ (i.e., Jeffreys divergence when written using the $\theta$-coordinate system).
This symmetrized divergence $\calS(P,Q)$ can be interpreted as the energy of the Riemannian length element $\ds$ along the primal geodesic $\gamma(t)$ and dual geodesic $\gamma^*(t)$ (with $\gamma(0)=\gamma^*(0)=P$ and $\gamma(1)=\gamma^*(1)=Q$), see~\cite{IG-2016}:
$\calS(P,Q)=\int_0^1 \ds^2(\gamma(t))\dt = \int_0^1 \ds^2(\gamma^*(t))\dt$. 
The Riemannian distance $\rho(P,Q)$ corresponds to the Riemannian length element along the Riemannian geodesic $\bar\gamma(t)$ induced by the Levi-Civita connection $\bar\nabla=\frac{\nabla+\nabla^*}{2}$: $\rho(P,Q)=\int_0^1 \ds(\bar\gamma(t))\dt$.

The inductive Gauss-Bregman center $C_\GB$ is obtained as a limit sequence of taking iteratively the $\nabla$-midpoints and $\nabla^*$-midpoints with respect to the $\nabla$ and $\nabla^*$ connections. Those midpoints correspond to the right and left centroids $C_{t+1}$ and $C_{t+1}^*$ with respect to $\calD(\cdot:\cdot)$:
\begin{eqnarray*}
C_{t+1} &=& \gamma_\nabla\left(C_t,C_t^*,\frac{1}{2}\right),\\
C_{t+1}^* &=& \gamma_{\nabla^*}\left(C_t,C_t^*,\frac{1}{2}\right),
\end{eqnarray*}
initialized with $\theta(C_0)=\sum_{i=1}^n w_i\theta(P_i)$ and $\eta(C_0^*)=\sum_{i=1}^n w_i\eta(P_i)$.
We have $C_0=\arg\min_{C\in M} \sum_i w_i \calD(P_i:C)$ and $C^*_0=\arg\min_{C\in M} \sum_i w_i \calD(P_i:C)$.
Figure~\ref{fig:GBDFS} illustrates geometrically the double sequence of taking iteratively dual geodesic midpoints to converge towards the Gauss-Bregman center $C_\GB$.
Thus the GB double sequence can be interpreted as a geometric optimization technique.

As a final remark, let us emphasize that choosing a proper mean or center depends on the application at hand~\cite{de2016mean,bullen2013handbook}.
For example, in Bayesian hypothesis testing, the Chernoff mean~\cite{chernoff1952measure} is used to upper bound Bayes' error, and has been widely used in information fusion~\cite{julier2017general} for its empricial robustness~\cite{nielsen2022revisiting} in practice. Jeffreys centroid has been successfully 
used in information retrieval tasks~\cite{Veldhuis-2002}.

\vskip 0.3cm
\noindent{\bf Code \& experiments}: The experiments were programmed in {\tt Java} (using the {\tt ApFloat} multiprecision-arithmetic package, \url{http://www.apfloat.org/}) and {\tt Maxima} (\url{https://maxima.sourceforge.io/}) and the various visualizations were programmed using the {\tt Processing} framework (\url{https://processing.org/}) and exported in Adobe PDF files.

\vskip 0.3cm
\noindent{\bf Acknowledgments.} The author warmly thanks Professor Kazuki Okamura (Shizuoka University, Japan) for numerous discussions on inductive means, specially for  potential extensions of Nakamura's inductive matrix geometric mean.

\bibliographystyle{plain}
\bibliography{JeffreysCentroidBIB}

\appendix

\section{Numerical Jeffreys centroids for categorical distributions}\label{sec:numericalJ}
Algorithm~\ref{algo:numJ} implements the method described in~\cite{nielsen2013jeffreys} for numerically finely approximating the Jeffreys centroid of a weighted set of categorical distributions.

\begin{algorithm}
\KwIn{A set of weighted categorical distributions: $\calP^w=\{p_1,\ldots,p_n\}$ with $w\in\Delta_n$ and $p_i\in\Delta_d$. 
Let $p_{i,j}$ denote the $j$-th component of $p_i$.}
\KwIn{A precision parameter $\eps>0$}
\KwOut{A numerical approximation of the SKL centroid/Jeffreys centroid $c$}
\tcc{Arithmetic weighted mean (normalized)}
$a_j=\sum_{i=1}^n w_ip_{i,j}$ for $i\in\{1,\ldots,d\}$ for $j\in\{1,\ldots,d\}$ \;
\tcc{Normalized geometric weighted mean}
$g_j=\frac{\prod_{i=1}^n p_{i,j}^{w_i}}{\sum_{j=1}^d \prod_{i=1}^n p_{i,j}^{w_i} }$ for $j\in\{1,\ldots,d\}$ \;
\tcc{Initialize range where to find the optimal $\lambda^*$}
$\lambda_M=0$; $\lambda_m=\max_{i\in\{1,\ldots,n\}}\{a_i+\log g_i\}-1$ \;
\tcc{Bisection search}
\While{$|\lambda_M-\lambda_m|>\eps$}{
$\lambda=\frac{\lambda_m+\lambda_M}{2}$ \;
\tcc{$W_0$ is the principal branch of Lambert $W$ function}
$c_j(\lambda)=\frac{a_j}{W_0\left(\frac{a_j}{g_j}\, e^{1+\lambda}\right)}$ for $i\in\{1,\ldots,d\}$ \;
\tcc{calculate the mass of $c(\lambda)$}
$s(\lambda)=\sum_{j=1}^d c_j$\;
\eIf{ $s(\lambda)>1$ }{\tcc{Consider next range $[\lambda,\lambda_M]$} $l_m=\lambda$ \; }{\tcc{Consider next range $[\lambda_m,\lambda]$}  $\lambda_M=\lambda$ \; }
}
$\lambda=\frac{\lambda_m+\lambda_M}{2}$ \;
$c_j(\lambda)=\frac{a_j}{W_0\left(\frac{a_j}{g_j}\, e^{1+\lambda}\right)}$ for $i\in\{1,\ldots,d\}$ \;
\Return $c(\lambda)$\;
\caption{Numerical approximation of the SKL/Jeffreys centroid for categorical distributions.}
\label{algo:numJ}
\end{algorithm}

\section{Closed-form formula for the symmetrized log det centroids}\label{sec:logdet}

Consider a set $\calP=\{P_1,\ldots, P_n\}$ of $n$ symmetric positive-definite matrices of the $d$-dimensional SPD cone $\Sym^{++}(d,\bbR)$ weighted by a vector $w=(w_1,\ldots,w_n)\in\Delta^\circ_n$ such that $P_i$ has weight $w_i$ for $i\in\{1,\ldots,n\}$.
The log det divergence~\cite{kulis2006learning} is a Bregman divergence induced by the strictly convex and differential generator $F_\ld(X)=-\log\det(X)$ on  $\Sym^{++}(d,\bbR)$ equipped with the inner product $\inner{X}{Y}=\tr(XY)$ for $X,Y\in\Sym(d,\bbR)$:
\begin{equation}
D_\ld(X:Y)=B_{F_\ld}(X:Y)=F(X)-F(Y)-\inner{X-Y}{\nabla F_\ld(Y)}.
\end{equation}

Since we have
$\nabla F_\ld(X)=-\nabla\log\det(X)=-\frac{\nabla\det(X)}{\det(X)}=-(X^{-1})^\top$ (hence $\nabla F_\ld(X)=-X^{-1}$ for symmetric matrices), it follows
that the log det divergence is:
\begin{eqnarray*}
D_\ld(X:Y) &=& \log\det(YX^{-1})+\tr((X-Y)Y^{-1}),\\
&=& \tr(XY^{-1})-\log\det(XY^{-1})-d,
\end{eqnarray*}
using the properties that $\det(X)\det(Y)=\det(XY)$, $\det(X^{-1})=\frac{1}{\det(X)}$ and $\tr(I)=d$ where $I$ denotes the $d\times d$ identity matrix.
When $d=1$, we recover the Itakura-Saito divergence~\cite{banerjee2005clustering} obtained for $F_\IS(x)=-\log x$ (Burg negative entropy) with $F_\IS'(x)=-\frac{1}{x}$:
$$
D_\IS(x:y)=B_{F_\IS}(x:y)=\frac{x}{y}-\log\frac{x}{y}-1, \quad x,y>0.
$$
The log det divergence  is known in statistics as Stein's loss~\cite{SteinLoss-1961,salehian2013recursive}, and has been used for estimating   covariance matrices.
the log det divergence $S_\ld$ satisfies the following invariance properties:
\begin{itemize}
\item Inversion invariance: $S_\ld(X^{-1},Y^{-1})=S_\ld(X,Y)$, and

\item Congruence invariance: For any invertible matrix $A\in\GL(d)$, we have $S_\ld(A X A^\top, A Y^{-1} A^\top)=S_\ld(X,Y)$.

\end{itemize}

The Jeffreys' symmetrized log det divergence (SLD) is thus:
\begin{eqnarray}
S_\ld(X,Y) &=& D_\ld(X:Y)+D_\ld(Y:X)=\tr\left(\left(Y^{-1}-X^{-1})(X-Y\right)\right),\\
&=& \tr\left(X^{-1}Y+Y^{-1}X-2I\right).
\end{eqnarray}

When $d=1$, the SLD corresponds to the COSH distance~\cite{gray1976distance} (COSine Hyperbolic distance,   the symmetrized Itakura-Saito divergence) when $d=1$:
$$
D_\COSH(x:y)=\left(\frac{y}{x}-\frac{1}{x}\right)=\frac{x}{y}+\frac{y}{x}-2.
$$

Consider a family $\calN_\mu=\left\{p_{\mu,\Sigma_1},\ldots, p_{\mu,\Sigma_n}\right\}$ of $n$ multivariate normal distributions centered at the same mean $\mu\in\bbR^d$ with covariance matrices $\Sigma_1,\ldots,\Sigma_n$.
The set of same-mean normal distributions forms an exponential family with natural parameter $\theta=\Sigma^{-1}$ (precision matrix) corresponding to the sufficient statistics $t(x)=-\frac{1}{2}xx^\top$, and log-normalizer $F(\theta)=-\frac{1}{2}\log\det(\theta)$.
Thus   the Kullback-Leibler divergence between $p_{\mu,\Sigma_i}$ and $p_{\mu,\Sigma_j}$  corresponds to a log det divergence~\cite{davis2006differential}:
$$
D_\KL[p_{\mu,\Sigma_i},p_{\mu,\Sigma_j}]=B_F(\theta_j:\theta_i)=D_\ld(\Sigma_j^{-1}:\Sigma_i^{-1}),
$$
and therefore the Jeffreys divergence $D_J[p_{\mu,\Sigma_i},p_{\mu,\Sigma_j}]$ corresponds to the matrix COSH/symetrized log-det divergence:
\begin{equation}
D_J[p_{\mu,\Sigma_i},p_{\mu,\Sigma_j}]=S_\ld(\Sigma_i^{-1},\Sigma_j^{-1})=\tr\left(\left(\Sigma_i^{-1}-\Sigma_j^{-1})(\Sigma_j-\Sigma_i\right)\right).
\end{equation}

The left KL centroid corresponds to a right Bregman centroid on the natural parameters (center of mass of the natural parameters) which corresponds to a weighted matrix harmonic mean on the covariance matrices:
$$
C_L^\KL=C_R^{B_F}=\left(\sum_{i=1}^n w_i \Sigma_i^{-1}\right)^{-1}.
$$

The right KL centroid is a left Bregman centroid (i.e., a quasi-arithmetic mean for $h(X)=-X^{-1}$ with $h^{-1}(Y)=-Y^{-1}$) which corresponds to the inverse of the weighted arithmetic mean on the covariance matrices:
$$
C_R^\KL=C_L^{B_F}=\left(\sum_{i=1}^n w_i \Sigma_i\right)^{-1}.
$$

We state the remarkable case of the closed-form formula for the symmetrized Bregman logdet centroid:

\begin{Proposition}[\cite{moakher2006symmetric}]
The symmetrized log det centroid of a set $\calP^w=\{(w_i,P_i)\}$ of $n$ weighted positive-definite matrices is $A\# H$ where $A=\sum_i w_iP_i$ and $H=\left(\sum_i w_i P_i^{-1}\right)^{-1}$ are the weighted arithmetic and harmonic means, and $A\# B$ is the matrix geometric mean.
\end{Proposition}

Since the proof was briefly sketched in~\cite{moakher2006symmetric}, we report it here in full length:
\begin{Proof}
We have 
$$
\min_{X} \sum_i w_i S_\ld(X,P_i) \equiv \min_X \tr\left(X^{-1}A+H^{-1}X\right).
$$
Setting the gradient of the right-hand side term to zero yields using matrix calculus~\cite{petersen2008matrix} yields:
\begin{eqnarray*}
\nabla_X \tr\left(X^{-1}A+H^{-1}X\right)= \tr\left( \nabla_X (X^{-1}A+H^{-1}X)\right)=0.
\end{eqnarray*}
Using the matrix calculus property that $\nabla (M^{-1})=-M^{-1}(\nabla M)M^{-1}$ for $M=X^{-1}A$, we get
$$
X^{-1}AX^{-1}-H^{-1}=0.
$$
That is, we need to solve the following Ricatti equation:
$$
X^{-1}AX^{-1}=H^{-1}.
$$
The well-known Ricatti equation $XA^{-1}X=B$ solves~\cite{bhatia2012riemannian} as $X=A\# B$, and therefore we get:
$$
X^{-1}=A^{-1}\# H^{-1}.
$$
Finally, we use the invariance property of the geometric mean under matrix inversion, $A^{-1}\# H^{-1}=A\# H$, to get the result 
$C^\ld_S=A\# H$.
\end{Proof}

The Riemannian Hessian metric $g(\theta)$ induced by $F(\theta)=-\frac{1}{2}\log\det(\theta)$ is
$$
g_\theta(S_1,S_2)=\tr\left(\theta^{-1} S_1 \theta^{-1} S_2\right),
$$ 
where $S_1$ and $S_2$ are two symmetric matrices of the tangent space $T_\theta$ at $\theta$.
The metric tensor $g$ is commonly called the trace metric or Affine-Invariant Riemannian Metric (AIRM)~\cite{thanwerdas2023n}.

It follows that the Riemannian geodesic midpoint is the matrix geometric mean~\cite{bhatia2006riemannian} given by
$$
X\# Y=X^{\frac{1}{2}}\, (X^{-\frac{1}{2}}\, Y\, X^{-\frac{1}{2}})^{\frac{1}{2}}\, X^{\frac{1}{2}}. 
$$

We have $\rho(X,X\# Y)=\rho(X\# Y,Y)$ where $\rho(\cdot,\cdot)$ denotes the geodesic length distance on the Riemannian manifold.
The geodesic length is given by the following formula~\cite{siegel1964symplectic,james1973variance}:
$$
\rho(P_1,P_2)=\left\| \log\left(P_1^{-\frac{1}{2}}\, P_2\, P_1^{-\frac{1}{2}}\right)\right\|_F  =\sqrt{\sum_{i=1}^d \log^2 \lambda_i\left(P_1^{-\frac{1}{2}}\, P_2\, P_1^{-\frac{1}{2}}\right)},
$$
where the $\lambda_i(X)$'s are the generalized eigenvalues of $X$.

We state the theorem characterizing geometrically the Jeffreys' centroid of a weighted set of centered multivariate normal distributions:

\begin{Theorem}[Jeffreys centroid of $n$ weighted centered normal distributions]\label{thm:normal}
The Jeffreys centroid $C_S$ of a weighted set $\{p_{\mu,\Sigma_i}\}$ of centered normal distributions $N(\mu,\Sigma_i)$ with weighted $w\in\Delta_n$ corresponds to the midpoint of the Fisher-Rao geodesic linking the left and right SKL centroids:
\begin{equation}
C_S  = \left(\sum_{i=1}^n w_i\Sigma_i\right)  \# \left(\sum_{i=1}^n w_i\Sigma_i^{-1}\right)^{-1},
\end{equation}
where $X\# Y$ is the geometric matrix mean:
$$ 
X\# Y=X^{\frac{1}{2}}\, (X^{-\frac{1}{2}}\, Y\, X^{-\frac{1}{2}})^{\frac{1}{2}}\, X^{\frac{1}{2}}.
$$
\end{Theorem}

This result first appeared in~\cite{moakher2006symmetric} (Lemma 17.4.3, item 3) and also appeared in an indirect but more general form in~\cite{SymMatISCentroid-2011} (Theorem 5.3).
Indeed, in~\cite{SymMatISCentroid-2011}, the authors define the regularized symmetric log det divergence as follows:
$$
S_\ld^\epsilon(X,Y)=\tr\left((X-Y)\left((Y+\eps I)^{-1}-(X+\eps I)^{-1}\right)\right),\quad \eps>0.
$$
This extended definition of the symmetrized logdet divergence allows one to consider degenerate semi-positive definite matrices.

\section{Fisher-Rao midpoint for multivariate normal distributions}\label{sec:frmvnmidpoint}

The expression of the Fisher-Rao geodesics for multivariate normal distributions passing through two given normal distributions was elucidated in~\cite{kobayashi2023geodesics}. We give below the recipe without the underlying geometric explanation that relies on a Riemannian submersion~\cite{kobayashi2023geodesics}.

\noindent\fbox{
\vbox{
\noindent\underline{Fisher-Rao geodesic midpoint $N=N(\mu,\Sigma)$ of $N_0=N(\mu_0,\Sigma_0)$ and $N_1=N(\mu_1,\Sigma_1)$}
\begin{itemize}

\item For $i\in\{0,1\}$, let $G_0=M_0\, D_0\, M_0^\top$ and $G_1=M_1\, D_1\, M_1^\top$, where 
\begin{eqnarray*}
D_0&=&\matthreethree{\Sigma^{-1}_0}{0}{0}{0}{1}{0}{0}{0}{\Sigma_0},\\
M_0&=&\matthreethree{I_d}{0}{0}{\mu_0^\top}{1}{0}{0}{-\mu_0}{I_d},\\
D_1&=&\matthreethree{\Sigma^{-1}_1}{0}{0}{0}{1}{0}{0}{0}{\Sigma_1},\\
M_1&=&\matthreethree{I_d}{0}{0}{\mu_1^\top}{1}{0}{0}{-\mu_1}{I_d},\\
\end{eqnarray*} 
where $I_d$ denotes the identity matrix of shape $d\times d$.
That is, matrices $G_0$ and $G_1\in\Sym_+(2d+1,\bbR)$ can be expressed by {block Cholesky factorizations}.

\item Consider the Riemannian geodesic midpoint $G$ in $\Sym_+(2d+1,\bbR)$ with respect to the trace metric:
$$
G=G_0^{\frac{1}{2}} \,\left(G_0^{-\frac{1}{2}}G_1G_0^{-\frac{1}{2}} \right)^{\frac{1}{2}}\, G_0^{\frac{1}{2}}.
$$

In order to compute the matrix power $G^p$ for $p\in\bbR$, we first calculate the Singular Value Decomposition 
 (SVD) of $G$: $G=O\, L\, O^\top$  (where $O$ is an orthogonal matrix and $L=\diag(\lambda_1,\ldots,\lambda_{2d+1})$ a diagonal matrix) and then get the matrix power as
$G^p=O\, L^p\, O^\top$ with $L^p=\diag(\lambda_1^p,\ldots, \lambda_{2d+1}^p)$.

\item  Retrieve $N=N(\mu,\Sigma)$ from matrix $G$:

\begin{eqnarray*}
\Sigma &=& [G]_{1:d,1:d}^{-1},\\
\mu&=& \Sigma\, [G]_{1:d,d+1},
\end{eqnarray*}
where
$[G]_{1:d,1:d}$ denotes the block matrix with rows and columns ranging from $1$ to $d$ extracted from $(2d+1)\times(2d+1)$ matrix $G$, and $[G]_{1:d,d+1}$ is similarly the column vector of $\bbR^d$ extracted from $G$.
\end{itemize}
}
}

\end{document}

%% file: macros-SBDCentroids.tex
\def\mattwotwo#1#2#3#4{{\left[\begin{array}{ll}#1 & #2\cr #3 & #4\end{array}\right]}}
\def\bbN{\mathbb{N}}
\def\matthreethree#1#2#3#4#5#6#7#8#9{{\left[\begin{array}{lll}#1 & #2 & #3
\cr #4 & #5 & #6\cr 
#7 & #8 & #9\end{array}\right]}}
\def\eqdef{:=}
\def\CO{\mathrm{CO}}
\def\ds{\mathrm{d}s}
\def\dt{\mathrm{d}t}
\def\diag{\mathrm{diag}}
\def\AGM{\mathrm{AGM}}
\def\JFR{\mathrm{JFR}}
\def\GB{\mathrm{GB}}

\def\du{\mathrm{d}u}

\def\eps{\epsilon}

\def\ld{{\mathrm{ld}}}
\def\bartheta{{\bar\theta}}
\def\ubartheta{{\underline{\theta}}}

\def\bbR{\mathbb{R}}
\def\bartheta{\bar\eta}

\def\calM{\mathcal{M}}
\def\st{\ :\ }
\def\calP{\mathcal{P}}
\def\calS{\mathcal{S}}
\def\calX{\mathcal{X}}
\def\calF{\mathcal{F}}
\def\KL{\mathrm{KL}}

\def\ds{\mathrm{d}s}

\def\tr{\mathrm{tr}}
\def\Cat{\mathrm{Cat}}
\def\Pr{\mathrm{Pr}}
\def\dmu{\mathrm{d}\mu}
\def\Sym{\mathrm{Sym}}
\def\IS{\mathrm{IS}}
\def\GL{\mathrm{GL}}
\def\COSH{\mathrm{COSH}}
\def\calN{\mathcal{N}}

\def\inner#1#2{{\langle #1,#2\rangle}}
\def\calE{\mathcal{E}}
\def\bartheta{{\bar\theta}}

\def\Cat{{\mathrm{Cat}}}

\def\inner#1#2{{\langle #1,#2\rangle}}
\def\dmu{\mathrm{d}\mu}
\def\bbR{\mathbb{R}}
\def\inner#1#2{{\langle #1,#2\rangle}}
\def\st{{\ :\ }}
\def\calP{\mathcal{P}}
\def\calM{\mathcal{M}}

\def\mattwotwo#1#2#3#4{{\left[\begin{array}{ll}#1 & #2\cr #3 & #4\end{array}\right]}}

\def\calD{\mathcal{D}}

%% file: JeffreysFisherRaoGaussBregmanCenters-arxivV1.bbl
\begin{thebibliography}{10}

\bibitem{almkvist1988gauss}
Gert Almkvist and Bruce Berndt.
\newblock {Gauss, Landen, Ramanujan, the arithmetic-geometric mean, ellipses,
  $\pi$, and the Ladies Diary}.
\newblock {\em The American Mathematical Monthly}, 95(7):585--608, 1988.

\bibitem{Amari-2007}
Shun-ichi Amari.
\newblock Integration of stochastic models by minimizing $\alpha$-divergence.
\newblock {\em Neural computation}, 19(10):2780--2796, 2007.

\bibitem{IG-2016}
Shun-ichi Amari.
\newblock {\em Information Geometry and Its Applications}.
\newblock Applied Mathematical Sciences. Springer Japan, 2016.

\bibitem{banerjee2005clustering}
Arindam Banerjee, Srujana Merugu, Inderjit~S Dhillon, Joydeep Ghosh, and John
  Lafferty.
\newblock {Clustering with Bregman divergences}.
\newblock {\em Journal of machine learning research}, 6(10), 2005.

\bibitem{barndorff2014information}
Ole Barndorff-Nielsen.
\newblock {\em Information and exponential families: in statistical theory}.
\newblock John Wiley \& Sons, 2014.

\bibitem{basseville2013divergence}
Mich{\'e}le Basseville.
\newblock {Divergence measures for statistical data processing: An annotated
  bibliography}.
\newblock {\em Signal Processing}, 93(4):621--633, 2013.

\bibitem{ben1989entropic}
Aharon Ben-Tal, Abraham Charnes, and Marc Teboulle.
\newblock Entropic means.
\newblock {\em Journal of Mathematical Analysis and Applications},
  139(2):537--551, 1989.

\bibitem{bhatia2012riemannian}
Rajendra Bhatia.
\newblock {The Riemannian mean of positive matrices}.
\newblock In {\em Matrix information geometry}, pages 35--51. Springer, 2012.

\bibitem{bhatia2006riemannian}
Rajendra Bhatia and John Holbrook.
\newblock Riemannian geometry and matrix geometric means.
\newblock {\em Linear algebra and its applications}, 413(2-3):594--618, 2006.

\bibitem{bullen2013handbook}
Peter~S Bullen.
\newblock {\em Handbook of means and their inequalities}, volume 560.
\newblock Springer Science \& Business Media, 2013.

\bibitem{bullen2003quasi}
Peter~S Bullen and PS~Bullen.
\newblock Quasi-arithmetic means.
\newblock {\em Handbook of means and their inequalities}, pages 266--320, 2003.

\bibitem{calvo1990distance}
Miquel Calvo and Josep~M Oller.
\newblock {A distance between multivariate normal distributions based in an
  embedding into the Siegel group}.
\newblock {\em Journal of multivariate analysis}, 35(2):223--242, 1990.

\bibitem{vcencov1978algebraic}
N.~N. {\v{C}}encov.
\newblock Algebraic foundation of mathematical statistics.
\newblock {\em Statistics: A Journal of Theoretical and Applied Statistics},
  9(2):267--276, 1978.

\bibitem{chandrasekhar2012compressed}
Vijay Chandrasekhar, Gabriel Takacs, David~M Chen, Sam~S Tsai, Yuriy Reznik,
  Radek Grzeszczuk, and Bernd Girod.
\newblock Compressed histogram of gradients: A low-bitrate descriptor.
\newblock {\em International journal of computer vision}, 96:384--399, 2012.

\bibitem{chernoff1952measure}
Herman Chernoff.
\newblock A measure of asymptotic efficiency for tests of a hypothesis based on
  the sum of observations.
\newblock {\em The Annals of Mathematical Statistics}, pages 493--507, 1952.

\bibitem{corless1996lambert}
Robert~M Corless, Gaston~H Gonnet, David~EG Hare, David~J Jeffrey, and Donald~E
  Knuth.
\newblock {On the Lambert W function}.
\newblock {\em Advances in Computational mathematics}, 5:329--359, 1996.

\bibitem{davis2006differential}
Jason Davis and Inderjit Dhillon.
\newblock Differential entropic clustering of multivariate gaussians.
\newblock {\em Advances in Neural Information Processing Systems}, 19, 2006.

\bibitem{de2016mean}
Miguel De~Carvalho.
\newblock Mean, what do you mean?
\newblock {\em The American Statistician}, 70(3):270--274, 2016.

\bibitem{fuglede2004jensen}
Bent Fuglede and Flemming Topsoe.
\newblock {Jensen-Shannon divergence and Hilbert space embedding}.
\newblock In {\em International symposium on Information theory (ISIT)},
  page~31. IEEE, 2004.

\bibitem{ge2022active}
Pengqiang Ge, Yiyang Chen, Guina Wang, and Guirong Weng.
\newblock {An active contour model driven by adaptive local pre-fitting energy
  function based on Jeffreys divergence for image segmentation}.
\newblock {\em Expert Systems with Applications}, 210:118493, 2022.

\bibitem{gray1976distance}
Augustine Gray and John Markel.
\newblock Distance measures for speech processing.
\newblock {\em IEEE Transactions on Acoustics, Speech, and Signal Processing},
  24(5):380--391, 1976.

\bibitem{james1973variance}
Alan~Treleven James.
\newblock The variance information manifold and the functions on it.
\newblock In {\em Multivariate Analysis--III}, pages 157--169. Elsevier, 1973.

\bibitem{SteinLoss-1961}
William James and Charles Stein.
\newblock Estimation with quadratic loss.
\newblock In {\em Proceedings of the fourth Berkeley symposium on mathematical
  statistics and probability}, volume~1, pages 361--379, 1961.

\bibitem{Jeffreys-1998}
Harold Jeffreys.
\newblock {\em The theory of probability}.
\newblock OuP Oxford, 1998.

\bibitem{johnson2001symmetrizing}
Don~H Johnson, Sinan Sinanovic, et~al.
\newblock {Symmetrizing the Kullback-Leibler distance}.
\newblock {\em IEEE Transactions on Information Theory}, 1(1):1--10, 2001.

\bibitem{julier2017general}
Simon Julier and Jeffrey~K Uhlmann.
\newblock General decentralized data fusion with covariance intersection.
\newblock In {\em Handbook of multisensor data fusion}, pages 339--364. CRC
  Press, 2017.

\bibitem{karcher1977riemannian}
Hermann Karcher.
\newblock Riemannian center of mass and mollifier smoothing.
\newblock {\em Communications on pure and applied mathematics}, 30(5):509--541,
  1977.

\bibitem{SymMatISCentroid-2011}
Sejong Kim, Jimmie Lawson, and Yongdo Lim.
\newblock The matrix geometric mean of parameterized, weighted arithmetic and
  harmonic means.
\newblock {\em Linear algebra and its applications}, 435(9):2114--2131, 2011.

\bibitem{kobayashi2023geodesics}
Shimpei Kobayashi.
\newblock {Geodesics of multivariate normal distributions and a Toda lattice
  type Lax pair}.
\newblock {\em Physica Scripta}, 98(11):115241, 2023.

\bibitem{kulis2006learning}
Brian Kulis, M{\'a}ty{\'a}s Sustik, and Inderjit Dhillon.
\newblock Learning low-rank kernel matrices.
\newblock In {\em Proceedings of the 23rd international conference on Machine
  learning}, pages 505--512, 2006.

\bibitem{lehmer1971compounding}
Derrick~H Lehmer.
\newblock On the compounding of certain means.
\newblock {\em Journal of Mathematical Analysis and Applications},
  36(1):183--200, 1971.

\bibitem{lin1991divergence}
Jianhua Lin.
\newblock {Divergence measures based on the Shannon entropy}.
\newblock {\em IEEE Transactions on Information theory}, 37(1):145--151, 1991.

\bibitem{liu2014distributed}
Qiang Liu and Alexander~T Ihler.
\newblock Distributed estimation, information loss and exponential families.
\newblock {\em Advances in neural information processing systems}, 27, 2014.

\bibitem{lloyd1982least}
Stuart Lloyd.
\newblock {Least squares quantization in PCM}.
\newblock {\em IEEE transactions on information theory}, 28(2):129--137, 1982.

\bibitem{miyamoto2024closed}
Henrique~K Miyamoto, F{\'a}bio~CC Meneghetti, Julianna Pinele, and Sueli~IR
  Costa.
\newblock {On closed-form expressions for the Fisher--Rao distance}.
\newblock {\em Information Geometry}, pages 1--44, 2024.

\bibitem{moakher2006symmetric}
Maher Moakher and Philipp~G Batchelor.
\newblock {Symmetric positive-definite matrices: From geometry to applications
  and visualization}.
\newblock In {\em Visualization and processing of tensor fields}, pages
  285--298. Springer, 2006.

\bibitem{murtagh2014ward}
Fionn Murtagh and Pierre Legendre.
\newblock {Ward’s hierarchical agglomerative clustering method: which
  algorithms implement Ward’s criterion?}
\newblock {\em Journal of classification}, 31:274--295, 2014.

\bibitem{nakamura2001algorithms}
Yoshimasa Nakamura.
\newblock Algorithms associated with arithmetic, geometric and harmonic means
  and integrable systems.
\newblock {\em Journal of computational and applied mathematics},
  131(1-2):161--174, 2001.

\bibitem{nielsen2013jeffreys}
Frank Nielsen.
\newblock Jeffreys centroids: A closed-form expression for positive histograms
  and a guaranteed tight approximation for frequency histograms.
\newblock {\em IEEE Signal Processing Letters}, 20(7):657--660, 2013.

\bibitem{nielsen2019jensen}
Frank Nielsen.
\newblock {On the Jensen--Shannon symmetrization of distances relying on
  abstract means}.
\newblock {\em Entropy}, 21(5):485, 2019.

\bibitem{nielsen2020generalization}
Frank Nielsen.
\newblock {On a generalization of the Jensen--Shannon divergence and the
  Jensen--Shannon centroid}.
\newblock {\em Entropy}, 22(2):221, 2020.

\bibitem{nielsen2022revisiting}
Frank Nielsen.
\newblock {Revisiting Chernoff information with likelihood ratio exponential
  families}.
\newblock {\em Entropy}, 24(10):1400, 2022.

\bibitem{nielsen2023simple}
Frank Nielsen.
\newblock {A simple approximation method for the Fisher--Rao distance between
  multivariate normal distributions}.
\newblock {\em Entropy}, 25(4):654, 2023.

\bibitem{InductiveMean-2023}
Frank Nielsen.
\newblock What is... an inductive mean?
\newblock {\em Notices of the American Mathematical Society},
  70(11):1851--1855, 2023.

\bibitem{NIELSEN2024}
Frank Nielsen.
\newblock {Approximation and bounding techniques for the Fisher-Rao distances
  between parametric statistical models}.
\newblock Handbook of Statistics. Elsevier, 2024.

\bibitem{nielsen2024divergences}
Frank Nielsen.
\newblock Divergences induced by the cumulant and partition functions of
  exponential families and their deformations induced by comparative convexity.
\newblock {\em Entropy}, 26(3):193, 2024.

\bibitem{nielsen2011burbea}
Frank Nielsen and Sylvain Boltz.
\newblock {The Burbea-Rao and Bhattacharyya centroids}.
\newblock {\em IEEE Transactions on Information Theory}, 57(8):5455--5466,
  2011.

\bibitem{nielsen2009sided}
Frank Nielsen and Richard Nock.
\newblock {Sided and symmetrized Bregman centroids}.
\newblock {\em IEEE transactions on Information Theory}, 55(6):2882--2904,
  2009.

\bibitem{nock2008mixed}
Richard Nock, Panu Luosto, and Jyrki Kivinen.
\newblock {Mixed Bregman clustering with approximation guarantees}.
\newblock In {\em Joint european conference on machine learning and knowledge
  discovery in databases}, pages 154--169. Springer, 2008.

\bibitem{petersen2008matrix}
Kaare~Brandt Petersen, Michael~Syskind Pedersen, et~al.
\newblock The matrix cookbook.
\newblock {\em Technical University of Denmark}, 7(15):510, 2008.

\bibitem{LegendreType-1967}
R.~T. Rockafellar.
\newblock {Conjugates and Legendre transforms of convex functions}.
\newblock {\em Canadian Journal of Mathematics}, 19:200--205, 1967.

\bibitem{salehian2013recursive}
Hesamoddin Salehian, Guang Cheng, Baba~C Vemuri, and Jeffrey Ho.
\newblock {Recursive estimation of the Stein center of SPD matrices and its
  applications}.
\newblock In {\em Proceedings of the IEEE International Conference on Computer
  Vision}, pages 1793--1800, 2013.

\bibitem{seal2020fuzzy}
Ayan Seal, Aditya Karlekar, Ondrej Krejcar, and Consuelo Gonzalo-Martin.
\newblock {Fuzzy $c$-means clustering using Jeffreys-divergence based
  similarity measure}.
\newblock {\em Applied Soft Computing}, 88:106016, 2020.

\bibitem{shima2007geometry}
Hirohiko Shima.
\newblock {\em {The geometry of Hessian structures}}.
\newblock World Scientific, 2007.

\bibitem{siegel1964symplectic}
CL~Siegel.
\newblock Symplectic geometry.
\newblock {\em Am. J. Math.}, 65:1--86, 1964.

\bibitem{Skovgaard-1984}
Lene~Theil Skovgaard.
\newblock {A Riemannian geometry of the multivariate normal model}.
\newblock {\em Scandinavian journal of statistics}, pages 211--223, 1984.

\bibitem{sra2021metrics}
Suvrit Sra.
\newblock {Metrics induced by Jensen-Shannon and related divergences on
  positive definite matrices}.
\newblock {\em Linear Algebra and its Applications}, 616:125--138, 2021.

\bibitem{sturm2003probability}
Karl-Theodor Sturm.
\newblock Probability measures on metric spaces of nonpositive.
\newblock {\em Heat kernels and analysis on manifolds, graphs, and metric
  spaces}, 338:357, 2003.

\bibitem{tabibian2015speech}
Shima Tabibian, Ahmad Akbari, and Babak Nasersharif.
\newblock Speech enhancement using a wavelet thresholding method based on
  symmetric kullback--leibler divergence.
\newblock {\em Signal Processing}, 106:184--197, 2015.

\bibitem{thanwerdas2023n}
Yann Thanwerdas and Xavier Pennec.
\newblock {$O(n)$-invariant Riemannian metrics on SPD matrices}.
\newblock {\em Linear Algebra and its Applications}, 661:163--201, 2023.

\bibitem{vajda2009metric}
Igor Vajda.
\newblock On metric divergences of probability measures.
\newblock {\em Kybernetika}, 45(6):885--900, 2009.

\bibitem{Veldhuis-2002}
Raymond Veldhuis.
\newblock {The centroid of the symmetrical Kullback-Leibler distance}.
\newblock {\em IEEE signal processing letters}, 9(3):96--99, 2002.

\bibitem{welk2006tensor}
Martin Welk, Christian Feddern, Bernhard Burgeth, and Joachim Weickert.
\newblock {Tensor median filtering and $M$-smoothing}.
\newblock {\em Visualization and Processing of Tensor Fields}, pages 345--356,
  2006.

\bibitem{zhao2014tensor}
Qibin Zhao, Guoxu Zhou, Liqing Zhang, and Andrzej Cichocki.
\newblock {Tensor-variate Gaussian processes regression and its application to
  video surveillance}.
\newblock In {\em 2014 IEEE International Conference on Acoustics, Speech and
  Signal Processing (ICASSP)}, pages 1265--1269. IEEE, 2014.

\end{thebibliography}
